\documentclass[journal, 10pt, onecolumn, letterpaper, oneside, final]{IEEEtran}

\usepackage{ifpdf}
\usepackage[noadjust]{cite}

\ifCLASSINFOpdf
  \usepackage[pdftex]{graphicx}
\else
  \usepackage[dvips]{graphicx}
\fi

\usepackage[cmex10]{amsmath}
\usepackage{array}

\ifCLASSOPTIONcompsoc
  \usepackage[caption=false,font=normalsize,labelfont=sf,textfont=sf]{subfig}
\else
  \usepackage[caption=false,font=footnotesize]{subfig}
\fi

\usepackage{fixltx2e}

\usepackage{url}

\usepackage[utf8]{inputenc} 
\usepackage[T1]{fontenc}
\usepackage{ifthen}

\usepackage{mathtools}
\usepackage{amssymb}
\usepackage{relsize}
\usepackage{bbm}
\usepackage{xfrac}
\usepackage{amsthm}
\theoremstyle{plain}
\newtheorem{definition}{Definition}
\newtheorem{lemma}{Lemma}

\newtheorem{theorem}{Theorem}

\newtheorem{remark}{Remark}

\newtheorem{example}{Example}
\newtheorem{proposition}{Proposition}
\usepackage[varg, cmbraces]{newtxmath}

\usepackage{xcolor}
\definecolor{burgundy}{rgb}{0.545098,0,0}
\definecolor{navyblue}{rgb}{0.0, 0.0, 0.5}
\definecolor{leafgreen}{rgb}{0.290196, 0.470588, 0.0}
\definecolor{bluegreen}{rgb}{0, 0.470588, 0.415686}
\definecolor{zuhl}{rgb}{0.1875, 0.26171875, 0.46484375}
\definecolor{orange}{rgb}{1, 0.6470588235, 0}
\definecolor{red}{rgb}{1, 0, 0}

\usepackage{color}
\usepackage{overpic}
\usepackage{pict2e}
\usepackage{booktabs}

\newcommand{\bvec}[1]{\boldsymbol{#1}}

\newcommand{\argmax}{\operatorname{arg~max}\limits}

\newcommand{\lemref}[1]{Lemma~\ref{#1}}
\newcommand{\propref}[1]{Proposition~\ref{#1}}
\newcommand{\thref}[1]{Theorem~\ref{#1}}

\newcommand{\sectref}[1]{Section~\ref{#1}}
\newcommand{\remref}[1]{Remark~\ref{#1}}

\newcommand{\appref}[1]{Appendix~\ref{#1}}

\usepackage{soul}
\setstcolor{red}

\allowdisplaybreaks[3]

\usepackage{array}
\usepackage[linesnumbered, algoruled, boxed, lined, vlined]{algorithm2e}
\SetKwRepeat{Do}{do}{while}

\usepackage{hyperref}
\hypersetup{
    colorlinks,
    linkcolor = {red!60!black},
    citecolor = {blue!60!black},
    urlcolor = {blue!50!black}
}
\usepackage{microtype}

\begin{document}

\title{Second- and Third-Order Asymptotics of the Continuous-Time Poisson Channel}

\IEEEoverridecommandlockouts

\author{%
\IEEEauthorblockN{%
Yuta~Sakai,~\IEEEmembership{Member,~IEEE,}
Vincent~Y.~F.~Tan,~\IEEEmembership{Senior~Member,~IEEE,} and
Mladen~Kova\v{c}evi\'{c},~\IEEEmembership{Member,~IEEE,}%
}%
\thanks{This work is supported by a Singapore National Research Foundation (NRF) Fellowship (R-263-000-D02-281), a Singapore Ministry of Education Tier 2 Grant (R-2630-000-C83-112), and the European Union's Horizon 2020 research and innovation programme under Grant Agreement number 856697.}
\thanks{Y.~Sakai is with the Department of Electrical and Computer Engineering, National University of Singapore, Singapore, Email: \url{eleyuta@nus.edu.sg}.
V.~Y.~F.~Tan is with the Department of Electrical and Computer Engineering and the Department of Mathematics, National University of Singapore, Singapore, Email: \url{vtan@nus.edu.sg}.
M.~Kova\v{c}evi\'{c} is with the Faculty of Technical Sciences, University of Novi Sad, Serbia, Email: \url{kmladen@uns.ac.rs}.}
\thanks{This paper was presented in part at the 2019 IEEE Information Theory Workshop (ITW2019), Visby, Gotland, Sweden \cite{sakai_kovacevic_tan_itw2019}.}
}%

\maketitle

\begin{abstract}
The paper derives the optimal second-order coding rate for the continuous-time Poisson channel.
We also obtain bounds on the third-order coding rate.
This is the first instance of a second-order result for a continuous-time channel.
The converse proof hinges on a novel construction of an output distribution induced by Wyner's discretized channel and the construction of an appropriate $\epsilon$-net of the input probability simplex.
While the achievability proof  follows the general program to prove the third-order term for non-singular discrete memoryless channels put forth by Polyanskiy, several non-standard techniques---such as new definitions and bounds on the probabilities of typical sets using logarithmic Sobolev inequalities---are employed to handle the continuous nature of the channel.
\end{abstract}


\IEEEpeerreviewmaketitle

\section{Introduction}

This study explores fundamental limits in Poisson communication theory \cite{verdu_1999} from the perspective of fixed-error probability asymptotics \cite{strassen_1962, hayashi_2009, polyanskiy_poor_verdu_2010, polyanskiy_thesis, tomamichel_tan_2013, tan_2014, kostina_verdu_2015, tan_tomamichel_2015, moulin_2017, kosut_sankar_2017}.
The continuous-time Poisson channel---simply referred to as the {\em Poisson channel}---is a canonical optical direct-detection communication model \cite{mazo_salz_1976, personick_1973}.
The output of the Poisson channel is a Poisson counting process whose intensity is determined by the sum of a \emph{dark current noise} and an \emph{input waveform} subject to peak and average power constraints; for more details refer to \sectref{sect:preliminaries} of this paper.
Kabanov \cite{kabanov_1978} derived the capacity,  i.e., the optimal first-order coding rate, of the Poisson channel in the absence of an average power constraint. 
Davis \cite{davis_1980} generalized Kabanov's capacity formula with an average power constraint.

While Kabanov's \cite{kabanov_1978} and Davis' \cite{davis_1980} proofs involved martingale techniques, Wyner \cite{wyner_1988} provided an alternative proof based on a discretization technique. 
In particular, Wyner discretized the Poisson channel into a binary-input binary-output discrete memoryless channel. He then applied elementary information-theoretic techniques to analyze the capacity of this  channel and related the capacity to that of the original Poisson channel.
Using the same discretization argument, Wyner \cite{wyner_1988} derived the error exponent \cite{gallager_1965} of the Poisson channel for all rates below capacity.

Recently, to understand the finite blocklength performance of channel coding, refinements of asymptotic estimates on optimal coding rates with fixed error probability have gained  increasing traction \cite{strassen_1962, hayashi_2009, polyanskiy_poor_verdu_2010, polyanskiy_thesis, tomamichel_tan_2013, tan_2014, tan_tomamichel_2015}.
In this paper, we derive the optimal second-order coding rate for the Poisson channel.
We also show an upper and a lower bounds on the optimal third-order coding rate.
This is the first instance of a conclusive second-order result in continuous-time communications in information theory.

While martingale-based techniques for Poisson channels enjoy several advantages~\cite{Shende_Wagner_2019}, our proof techniques are inspired by Wyner's discretization argument   \cite{wyner_1988}.
In the converse part, particularly in the application of the meta-converse~\cite{polyanskiy_poor_verdu_2010} to the discretized channel, we construct a somewhat artificial output distribution induced by Wyner's discretized channel and the construction of an appropriate $\epsilon$-net of the input probability simplex.
This construction differs from existing constructions in the literature~\cite{hayashi_2009, tomamichel_tan_2013, kostina_verdu_2015} and appears to be essential in handling the continuous nature of the channel model.
In the achievability part, we adapt Polyanskiy's technique to prove the third-order asymptotics for non-singular%
\footnote{A channel $W : \mathcal{X} \to \mathcal{Y}$ is said to be \emph{singular} if $W(y \mid x) \, W(y \mid z) > 0$ implies that $W(y \mid x) = W(y \mid z)$ (cf.\ \cite[Section~4.2.1]{tan_2014}).
A channel is said to be \emph{non-singular} if it is not singular.}
discrete memoryless channels (DMCs)~\cite[Section~3.4.5]{polyanskiy_thesis} to our scenario.
A straightforward generalization of the random coding union bound \cite[Theorem~16]{polyanskiy_poor_verdu_2010} and a judicious choice of the input distribution inspired by the delta-convention  \cite[Convention~2.11]{csiszar_korner_2011}  are proposed in view of the additive cost constraint imposed on the discretized channel. 
In addition, to bound a certain probability that results from the continuous nature of the Poisson channel and certain properties of its discretized version, a modified logarithmic Sobolev inequality (cf.\ \cite[Chapter~6]{boucheron_lugosi_massart_2013}) rather than Hoeffding's inequality is employed.

The rest of this paper is organized as follows:
\sectref{sect:preliminaries} introduces the formal definition of the Poisson channel, its classical channel capacity \cite{kabanov_1978, davis_1980, wyner_1988} is stated in \thref{th:1st-order}, and our second-order asymptotics for the Poisson channel---i.e., the main result of this study---is presented in \thref{th:2nd-order} together with certain bounds on the third-order coding rate.
\sectref{sect:main} describes the proof of \thref{th:2nd-order}.
This section is partitioned into various subsections. 
In \sectref{sect:notations}, we introduce the basic notations in the the study of second- and third-order asymptotics.
In \sectref{sect:discretization}, we revisit Wyner's discretization argument \cite{wyner_1988}.
The proofs of the converse and achievability parts of \thref{th:2nd-order} are provided in Sections~\ref{sect:converse} and~\ref{sect:direct}, respectively.
\sectref{sect:conclusion} discusses technical novelties and contributions of our proofs;
readers may benefit from reading through the discussion in \sectref{sect:conclusion} before perusing the proofs.

\section{Continuous-Time Poisson Channel and Its Second- and Third-Order Asymptotics}
\label{sect:preliminaries}

We now introduce the mathematical model of the \emph{continuous-time Poisson channel}. 
An input of the Poisson channel is an integrable function $\lambda : [0, \infty) \to [0, \infty)$ called a \emph{waveform.}
Given a waveform $\lambda$ and a constant $\lambda_{0} \ge 0$ called the \emph{dark current,} the output of the Poisson channel is modeled by a Poisson counting process $\{ \nu( t ) \}_{t \ge 0}$ of intensity $\lambda( t ) + \lambda_{0}$, i.e.,
\begin{align}
\mathbb{P}\{ \nu( 0 ) = 0 \}
& =
1 ,
\\
\mathbb{P}\{ \nu( t+\tau ) - \nu( t ) = k \}
& =
\frac{ \mathrm{e}^{-\Lambda} \Lambda^{k} }{ k! }
\end{align}
for every $0 \le t, \tau < \infty$ and every $k \in \mathbb{N} \cup \{ 0 \}$, where
\begin{align}
\Lambda
=
\Lambda( \lambda, \lambda_{0}, t, \tau )
\coloneqq
\int_{t}^{t + \tau} \Big( \lambda(u) + \lambda_{0} \Big) \, \mathrm{d} u .
\end{align}
Here, the Poisson counting process $\{ \nu( t ) \}_{t \ge 0}$ is defined as a random mapping $\nu : [0, \infty) \to \mathbb{N} \cup \{ 0 \}$.
For the sake of brevity, we denote  the random mapping $\nu : [0, T] \to \mathbb{N} \cup \{ 0 \}$ as $\nu_{0}^{T}$.

Let $T > 0$, $A > 0$, and $0 < \sigma \le 1$ be three constants.
Denote by $\mathcal{S}( T )$ the collection of non-decreasing   functions $g : [0, T] \to \mathbb{N} \cup \{ 0 \}$ satisfying $g( 0 ) = 0$.
In addition, denote by $\mathcal{W}(T, A, \sigma)$ the collection of waveforms $\lambda : [0, T] \to [0, A]$ satisfying the average power constraint
\begin{align}
\frac{ 1 }{ T } \int_{0}^{T} \lambda( t ) \, \mathrm{d} t
\le
\sigma A .
\label{eq:average_power_constraint}
\end{align}
Then, a channel code for the Poisson channel with dark current $\lambda_{0}$ can be defined as follows:

\begin{definition}
\label{def:code_Poisson}
Given an integer $M \ge 1$ and a real number $0 \le \varepsilon \le 1$, a pair of encoder $\phi : \{ 1, \dots, M \} \to \mathcal{W}(T, A, \sigma)$ and decoder $\psi : \mathcal{S}( T ) \to \{ 1, \dots, M \}$ is called a \emph{$(T, M, A, \sigma, \varepsilon)_{\mathrm{avg}}$-code} if
\begin{align}
\frac{ 1 }{ M } \sum_{m = 1}^{M} \mathbb{P}\{ \psi( \nu_{0}^{T} ) = m \mid \lambda = \phi( m ) \}
\ge
1 - \varepsilon .
\end{align}
Here, $\lambda$ stands for the $\mathcal{W}(T, A, \sigma)$-valued random variable (r.v.) induced by the encoder $\phi$ and the uniformly distributed messages on $\{ 1, \dots, M \}$.
\end{definition}

For $0 < \varepsilon < 1$, denote by $M_{\mathrm{avg}}^{\ast}(\lambda_{0}, T, A, \sigma, \varepsilon)$ the maximum  integer $M$ such that a $(T, M, A, \sigma, \varepsilon)_{\mathrm{avg}}$-code exists for the Poisson channel with dark current $\lambda_{0}$.
Assume throughout this paper that all logarithms are the natural logarithm.

\begin{theorem}[{\cite{kabanov_1978, davis_1980, wyner_1988}}]
\label{th:1st-order}
It holds that
\begin{align}
\log M_{\mathrm{avg}}^{\ast}(\lambda_{0}, T, A, \sigma, \varepsilon)
=
T \, C^{\ast} + \mathrm{o}( T )
\qquad
(\mathrm{as} \ T \to \infty) ,
\label{eq:1st_order}
\end{align}
where the Poisson channel capacity $C^{\ast}$ is given by
\begin{align}
C^{\ast}
& \: =
C^{\ast}(\lambda_{0}, A, \sigma)
\notag \\
& \coloneqq
A \, \bigg( (1-p^{\ast}) \, s \log \frac{ s }{ p^{\ast}+s } + p^{\ast} \, (1+s) \log \frac{ 1+s }{ p^{\ast}+s } \bigg) ,
\label{def:capacity_Poisson}
\end{align}
and three numbers $s$, $p^{\ast}$, and $p_{0}$ are given by
\begin{align}
s
=
s(\lambda_{0}, A)
& \coloneqq
\frac{ \lambda_{0} }{ A } ,
\label{def:s} \\
p^{\ast}
=
p^{\ast}(\lambda_{0}, A, \sigma)
& \coloneqq
\min\{ \sigma, p_{0} \} ,
\label{def:p_ast} \\
p_{0}
=
p_{0}(\lambda_{0}, A)
& \coloneqq
\frac{ (1+s)^{1+s} }{ s^{s} \, \mathrm{e} } - s .
\label{def:p0}
\end{align}
\end{theorem}

Our goal is to refine the $+ \mathrm{o}( T )$ term in \eqref{eq:1st_order}. We do so via Wyner's discretization argument \cite[Section~II in Part~I]{wyner_1988} and finite blocklength analyses.
The following theorem exactly characterizes the optimal second-order coding rate of the Poisson channel together with bounds on the third-order coding rate.

\begin{theorem}
\label{th:2nd-order}
It holds that
\begin{align}
\log M_{\mathrm{avg}}^{\ast}(\lambda_{0}, T, A, \sigma, \varepsilon)
=
T \, C^{\ast} + \sqrt{ T \, V^{\ast} } \, \Phi^{-1}( \varepsilon ) + \rho_{T} ,
\end{align}
where $\rho_{T} = \mathrm{O}( \log T )$ as $T \to \infty$, the Poisson channel dispersion $V^{\ast}$ is given by
\begin{align}
V^{\ast}
& \: =
V^{\ast}(\lambda_{0}, A, \sigma)
\notag \\
& \coloneqq
A \, \bigg( (1-p^{\ast}) \, s \log^{2} \frac{ s }{ p^{\ast}+s } + p^{\ast} \, (1+s) \log^{2} \frac{ 1+s }{ p^{\ast}+s } \bigg) ,
\label{def:dispersion_Poisson}
\end{align}
and $\Phi^{-1}( \cdot )$ stands for the inverse function of the standard Gaussian cumulative distribution function
\begin{align}
\Phi( u )
\coloneqq
\frac{ 1 }{ \sqrt{ 2 \pi } } \int_{-\infty}^{u} \mathrm{e}^{-t^{2}/2} \, \mathrm{d}t .
\end{align}
More precisely, there exist positive constants $c_{1}$ and $c_{2}$ satisfying
\begin{align}
\frac{ 1 }{ 2 } \log T - c_{1}
\le
\rho_{T}
\le
\log T + c_{2}
\label{eq:third-order_bounds}
\end{align}
for sufficiently large $T$.
\end{theorem}

The converse and achievability parts of \thref{th:2nd-order} are proved in Sections~\ref{sect:converse} and~\ref{sect:direct}, respectively.
The technical contributions and novelties in the proofs of the converse and achievability parts will be discussed in detail in Sections~\ref{sect:comment_converse} and~\ref{sect:comment_direct}, respectively.

\section{Proof of \thref{th:2nd-order}}
\label{sect:main}

\subsection{Notations for Second-Order Asymptotic Analysis}
\label{sect:notations}

Given two discrete distributions $P$ and $Q$ on the same space,   define the following four divergences:
\begin{align}
D_{\mathrm{s}}^{\epsilon}(P \, \| \, Q)
& \coloneqq
\sup\bigg\{ R \in \mathbb{R} \ \bigg| \ \mathbb{P}\bigg\{ \log \frac{ P(Z) }{ Q(Z) } \le R \bigg\} \le \epsilon \bigg\} ,
\\
D(P \, \| \, Q)
& \coloneqq
\mathbb{E}\bigg[ \log \frac{ P(Z) }{ Q(Z) } \bigg] ,
\\
V(P \, \| \, Q)
& \coloneqq
\mathbb{E}\bigg[ \bigg( \log \frac{ P(Z) }{ Q(Z) } - D(P \, \| \, Q) \bigg)^{2} \bigg] ,
\\
\Xi(P \, \| \, Q)
& \coloneqq
\mathbb{E}\bigg[ \bigg| \log \frac{ P(Z) }{ Q(Z) } - D(P \, \| \, Q) \bigg|^{3} \bigg] ,
\end{align}
where $Z$ is a r.v.\ satisfying%
\footnote{The notation $\mathbb{P} \circ Z^{-1}$ stands for the probability distribution induced by the r.v.\ $Z$.}
$\mathbb{P} \circ Z^{-1} = P$.
Moreover, given countable alphabets $\mathcal{X}$ and $\mathcal{Y}$, a distribution $P$ on $\mathcal{X}$, a channel $W : \mathcal{X} \to \mathcal{Y}$, and a distribution $Q$ on $\mathcal{Y}$, define the following three conditional divergences:
\begin{align}
D(W \, \| \, Q \mid P)
& \coloneqq
\mathbb{E}\bigg[ \log \frac{ W(Y \mid X) }{ Q( Y ) } \bigg] ,
\\
V(W \, \| \, Q \mid P)
& \coloneqq
\mathbb{E}\Bigg[ \bigg( \log \frac{ W(Y \mid X) }{ Q( Y ) } - \mathbb{E}\bigg[ \log \frac{ W(Y \mid X) }{ Q( Y ) } \ \bigg| \ X \bigg] \bigg)^{2} \Bigg] ,
\\
\Xi(W \, \| \, Q \mid P)
& \coloneqq
\mathbb{E}\Bigg[ \bigg| \log \frac{ W(Y \mid X) }{ Q( Y ) } - \mathbb{E}\bigg[ \log \frac{ W(Y \mid X) }{ Q( Y ) } \ \bigg| \ X \bigg] \bigg|^{3} \Bigg] ,
\end{align}
where $(X, Y)$ is a pair of r.v.'s satisfying $\mathbb{P} \circ (X, Y)^{-1} = P \times W$, and $P \times W$ stands for the joint distribution on $\mathcal{X} \times \mathcal{Y}$ defined as
\begin{align}
(P \times W)(x, y)
\coloneqq
P(x) \, W(y \mid x)
\end{align}
for each $(x, y) \in \mathcal{X} \times \mathcal{Y}$.
In particular, we write
\begin{align}
I(P, W)
& \coloneqq
D(W \, \| \, PW \mid P)
=
D(P \times W \, \| \, P \times PW) ,
\\
V(P, W)
& \coloneqq
V(W \, \| \, PW \mid P) ,
\\
\tilde{V}(P, W)
& \coloneqq
V(P \times W \, \| \, P \times PW) ,
\\
\Xi(P, W)
& \coloneqq
\Xi(W \, \| \, PW \mid P) ,
\\
\tilde{\Xi}(P, W)
& \coloneqq
\Xi(P \times W \, \| \, P \times PW)
\end{align}
for a distribution $P$ on $\mathcal{X}$ and a channel $W : \mathcal{X} \to \mathcal{Y}$, where $PW$ stands for the output distribution on $\mathcal{Y}$ defined as
\begin{align}
PW( y )
\coloneqq
\sum_{x^{\prime} \in \mathcal{X}} (P \times W)(x^{\prime}, y)
\end{align}
for each $y \in \mathcal{Y}$.

\subsection{Discretization of the Poisson Channel}
\label{sect:discretization}

Our analyses hinge on  Wyner's \emph{ad hoc} assumption \cite[Section~II in Part I]{wyner_1988}. At a high level, this assumption says that the performance of the original channel is roughly equivalent to a discretized version of the  Poisson channel. We now introduce the discretized version of the Poisson channel as follows:
For $\Delta > 0$, define the binary asymmetric channel $W_{\Delta} : \{ 0, 1 \} \to \{ 0, 1 \}$ by%
\footnote{If the dark current of the Poisson channel is zero, i.e., $\lambda_{0} = 0$, then the channel $W_{\Delta} : \{ 0, 1 \} \to \{ 0, 1 \}$ is a Z-channel.}
\begin{align}
W_{\Delta}(1 \mid x)
\coloneqq
\begin{cases}
s \, A \, \Delta \, \mathrm{e}^{- s A \Delta}
& \mathrm{if} \ x = 0 ,
\\
(1+s) \, A \, \Delta \, \mathrm{e}^{- (1+s) A \Delta}
& \mathrm{if} \ x = 1 ,
\end{cases}
\label{def:W_Delta}
\end{align}
where $s$ is defined in \eqref{def:s}.
In particular, we define
\begin{align}
W_{n}
\coloneqq
W_{\Delta_{n}}
\end{align}
for each $n \ge 1$, where the number $\Delta_{n} > 0$ is given by
\begin{align}
\Delta_{n}
\coloneqq
\frac{ T }{ n } .
\label{def:Delta_n}
\end{align}
It is clear that $W_{n}(\cdot \mid 0)$ and $W_{n}(\cdot \mid 1)$ are Bernoulli distributions.
We write these Bernoulli parameters as
\begin{align}
a_{n}
& \coloneqq
W_{n}(1 \mid 0) ,
\label{def:an} \\
b_{n}
& \coloneqq
W_{n}(1 \mid 1) .
\label{def:bn}
\end{align}

Denote by $\mathcal{B}( n, \sigma )$ the set of $n$-length binary sequences $\bvec{x}  = (x_{1}, \dots, x_{n}) \in \{ 0, 1 \}^{n}$ satisfying the weight constraint
\begin{align}
\frac{ 1 }{ n } \sum_{i = 1}^{n} x_{i}
\eqqcolon
P_{\bvec{x}}( 1 )
\le
\sigma ,
\label{eq:weight_constraint}
\end{align}
where $P_{\bvec{x}}$ stands for the type or empirical distribution of the binary sequence $\bvec{x} \in \{ 0, 1 \}^{n}$ (see \cite{csiszar_korner_2011}).
We define a channel code for the discretized channel $W_{n}^{n}$ under the weight constraint $\sigma$ as follows:

\begin{definition}
Given an integer $M \ge 1$ and a real $0 \le \varepsilon \le 1$, a pair of encoder $\phi : \{ 1, \dots, M \} \to \mathcal{B}(n, \sigma)$ and decoder $\psi : \{ 0, 1 \}^{n} \to \{ 1, \dots, M \}$ is called an \emph{$(n, M, \sigma, \varepsilon)_{\mathrm{avg}}$-code for the discretized channel} $W_{n}^{n}$ if
\begin{align}
\frac{ 1 }{ M } \sum_{m = 1}^{M} W_{n}^{n}(\psi^{-1}( m ) \mid \phi( m ))
\ge
1 - \varepsilon ,
\end{align}
where $W^{n} : \mathcal{X}^{n} \to \mathcal{Y}^{n}$ stands for the $n$-fold product channel of a DMC $W : \mathcal{X} \to \mathcal{Y}$, i.e.,
\begin{align}
W^{n}(\bvec{y} \mid \bvec{x})
\coloneqq
\prod_{i=1}^{n} W(y_{i} \mid x_{i})
\end{align}
for each $\bvec{x} = (x_{1}, \dots, x_{n}) \in \mathcal{X}^{n}$ and $\bvec{y} = (y_{1}, \dots, y_{n}) \in \mathcal{Y}^{n}$.
\end{definition}

Wyner's \emph{ad hoc} assumption \cite[Section~II in Part~I]{wyner_1988} constrains a $(T, M, A, \sigma, \varepsilon)_{\mathrm{avg}}$-code for the Poisson channel in a certain way so that the resultant channel code is equivalent---in a sense to be made precise in~\lemref{lem:eps_nk}---to an $(n, M, \sigma, \varepsilon)_{\mathrm{avg}}$-code for the discretized channel $W_{n}^{n}$.
In fact, the Poisson channel and its discretized channel $W_{n}^{n}$ can be compared via a certain channel ordering, as shown in the following proposition.

\begin{proposition}
\label{prop:Shannon_ordering}
The Poisson channel is better in the Shannon sense than its discretized channel $W_{n}^{n}$, where we say that a channel $V_{1}$ is \emph{better in the Shannon sense than} another channel $V_{2}$ if for each $n \ge 1$ and each code for $V_{2}^{n}$, say $\mathcal{C}_2$,  there exists a code for $V_{1}^{n}$, say $\mathcal{C}_1$, with the same message size and the average probability of error of $\mathcal{C}_2$  is no larger than that for $\mathcal{C}_1$ (cf.\ \cite[Problem~6.17(a)]{csiszar_korner_2011}, \cite{shannon_1958}, \cite{nasser_2018}).
\end{proposition}

\begin{IEEEproof}[Proof of \propref{prop:Shannon_ordering}]
See \appref{app:Shannon_ordering}.
\end{IEEEproof}

Therefore, it is clear that a $(T, M, A, \sigma, \varepsilon)_{\mathrm{avg}}$-code exists for the Poisson channel, provided that an $(n, M, \sigma, \varepsilon)_{\mathrm{avg}}$-code exists for the discretized channel $W_{n}^{n}$.
This implies that
\begin{align}
M_{n}^{\ast}(\sigma, \varepsilon)
\le
M_{\mathrm{avg}}^{\ast}(\lambda_{0}, T, A, \sigma, \varepsilon) ,
\label{eq:comparison}
\end{align}
where $M_{n}^{\ast}(\sigma, \varepsilon)$ stands for the maximum integer $M$ such that an $(n, M, \sigma, \varepsilon)_\mathrm{avg}$-code exists for the discretized channel $W_{n}^{n}$.
Moreover, Wyner \cite[Theorem~2.1 in Part~II]{wyner_1988} showed that this discretization error is negligible for $n$ large enough.
The following lemma is a direct consequence of \cite[Theorem~2.1 in Part~II]{wyner_1988}.

\begin{lemma}
\label{lem:eps_nk}
There exist a sequence $\epsilon_{n} = \mathrm{o}( 1 )$ satisfying $0 < \epsilon_{n} < 1-\varepsilon$ and a subsequence $\{ n_{k} \}_{k = 1}^{\infty}$ such that
\begin{align}
M_{n_{k}}^{\ast}(\sigma, \varepsilon + \epsilon_{n_{k}})
=
M_{\mathrm{avg}}^{\ast}(\lambda_{0}, T, A, \sigma, \varepsilon)
\end{align}
for every $k \ge 1$.
\end{lemma}

\subsection{Proof of Converse Part of \thref{th:2nd-order}}
\label{sect:converse}

It follows from \lemref{lem:eps_nk} that
\begin{align}
\log M_{\mathrm{avg}}^{\ast}(\lambda_{0}, T, A, \sigma, \varepsilon)
\le
\limsup_{n \to \infty} \log M_{n}^{\ast}(\sigma, \varepsilon + \epsilon_{n}) ,
\end{align}
therefore, it suffices to prove the following lemma to assert the converse part of \thref{th:2nd-order}.

\begin{lemma}
\label{lem:converse}
For every $0 < \varepsilon < 1$, it holds that
\begin{align}
\limsup_{n \to \infty} \log M_{n}^{\ast}( \sigma, \varepsilon + \epsilon_{n} )
\le
T \, C^{\ast} + \sqrt{ T \, V^{\ast} } \, \Phi^{-1}( \varepsilon ) + \log T + \mathrm{O}( 1 )
\label{eq:converse}
\end{align}
as $T \to \infty$.
\end{lemma}

\begin{IEEEproof}[Proof of \lemref{lem:converse}]
Let $P^{\ast}$ be the capacity-achieving input distribution (CAID) for the Poisson channel, i.e., it is the Bernoulli distribution with parameter $p^{\ast}$, where $p^{\ast}$ is defined in \eqref{def:p_ast}.
By the definitions of $p^{\ast}$ and $p_{0}$ in \eqref{def:p_ast} and \eqref{def:p0}, respectively, we see that
\begin{align}
\min\bigg\{ \sigma, \frac{ 1 }{ \mathrm{e} } \bigg\}
\le
p^{\ast}
\le
\min\bigg\{ \sigma, \frac{ 1 }{ 2 } \bigg\} .
\label{eq:bounds_p_ast}
\end{align}
For each $- p^{\ast} \le u \le 1 - p^{\ast}$, we define the $u$-shifted distribution $P_{[u]}^{\ast}$ from the CAID $P^{\ast}$ as the Bernoulli distribution
\begin{align}
P_{[u]}^{\ast}( x )
=
\begin{cases}
1 - (p^{\ast} + u)
& \mathrm{if} \ x = 0 ,
\\
p^{\ast} + u
& \mathrm{if} \ x = 1 .
\end{cases}
\label{def:u-shift}
\end{align}
Now, construct the distribution $Q^{(n)}$ on $\{ 0, 1 \}^{n}$ as follows:%
\footnote{Throughout the proof of \lemref{lem:converse}, we assume that $T$ is large enough so that there exists at least one $m$ satisfying the condition of the third sum in \eqref{eq:artifical}.}
\begin{align}
Q^{(n)}( \bvec{y} )
\coloneqq
\frac{ 1 }{ 3 } \prod_{i = 1}^{n} P_{[-\kappa]}^{\ast}W_{n}( y_{i} ) + \frac{ 1 }{ 3 } \prod_{i = 1}^{n} P_{[\kappa]}^{\ast}W_{n}( y_{i} ) + \frac{ 1 }{ 3 \, F } \sum_{\substack{ m = -\infty : \\ 0 \le p^{\ast} + m/T \le 1 }}^{\infty} \mathrm{e}^{-\gamma m^{2}/T} \prod_{i = 1}^{n} P_{[m/T]}^{\ast}W_{n}( y_{i} )
\label{eq:artifical}
\end{align}
for each $n \ge 1$ and $\bvec{y} = (y_{1}, \dots, y_{n}) \in \{ 0, 1 \}^{n}$, where the constant $\kappa > 0$ is given by
\begin{align}
\kappa
\coloneqq
\frac{ 1 }{ 2 } \min\bigg\{ \sigma, \frac{ 1 }{ \mathrm{e} } \bigg\} ,
\label{def:kappa}
\end{align}
the constant $\gamma > 0$ will be specified later (specifically in \eqref{eq:choice_of_gamma} of \sectref{sect:I3}), and $F$ is the normalization constant ensuring that
\begin{align}
\sum_{\bvec{y} \in \{ 0, 1 \}^{n}} Q^{(n)}( \bvec{y} )
=
1 .
\end{align}
Note that the third component of $Q^{(n)}$ is a convex combination of a set of exponentially-weighted distributions indexed by an appropriately constructed $\epsilon$-net in the input probability simplex.
Also, we note that $F$ is positive, and it follows by the Gaussian integral that
\begin{align}
F
<
\sum_{m = -\infty}^{\infty} \mathrm{e}^{-\gamma m^{2}/T}
<
1 + \int_{-\infty}^{\infty} \mathrm{e}^{-\gamma m^{2}/T} \, \mathrm{d} m
=
1 + \sqrt{ \frac{ \pi \, T }{ \gamma } } .
\label{eq:F_bound}
\end{align}

Let $\mathbb{N} \coloneqq \{ 1, 2, \dots \}$ be the set of positive integers, and $\eta$ chosen so that%
\footnote{The maximum value of $\{ \epsilon_{n} \}_{n=1}^{\infty}$ exists, because $\epsilon_{n} = \mathrm{o}(1)$ as $n \to \infty$.}
\begin{align}
0
<
\eta
<
1 - \varepsilon - \max_{n \in \mathbb{N}} \epsilon_{n} ,
\end{align}
where $\{ \epsilon_{n} \}_{n = 1}^{\infty}$ is given in \lemref{lem:eps_nk}.
For each $n \ge 1$, choose a sequence $\bvec{x}^{(n)} = (x_{1}^{(n)}, \dots, x_{n}^{(n)}) \in \{ 0, 1 \}^{n}$ satisfying
\begin{align}
\bvec{x}^{(n)}
\in
\argmax\limits_{\bvec{x} \in \{ 0, 1 \}^{n} : P_{\bvec{x}}( 1 ) \le \sigma} D_{\mathrm{s}}^{\varepsilon + \epsilon_{n} + \eta}(W_{n}^{n}(\cdot \mid \bvec{x}) \, \| \, Q^{(n)}) .
\label{def:xn}
\end{align}
For short, we write
\begin{align}
P_{n}
& \coloneqq
P_{\bvec{x}^{(n)}} ,
\label{def:P_xn} \\
p_{n}
& \coloneqq
P_{n}( 1 ) ,
\label{def:pn} \\
r_{n}
& \coloneqq
P_{n}W_{n}( 1 ) .
\label{def:rn_converse}
\end{align}
Now consider the partition $\{ \mathcal{I}_{1}, \mathcal{I}_{2}, \mathcal{I}_{3} \}$ of $\mathbb{N}$ given by
\begin{align}
\mathcal{I}_{1}
& \coloneqq
\{ n \in \mathbb{N} \mid p_{n} \ge p^{\ast} + \kappa \} ,
\\
\mathcal{I}_{2}
& \coloneqq
\{ n \in \mathbb{N} \mid p_{n} \le p^{\ast} - \kappa \} ,
\\
\mathcal{I}_{3}
& \coloneqq
\{ n \in \mathbb{N} \mid |p_{n} - p^{\ast}| < \kappa \} .
\label{eq:I3}
\end{align}
Clearly, at least one of the subsets $\mathcal{I}_{1}, \mathcal{I}_{2}, \mathcal{I}_{3} \subset \mathbb{N}$ must be countably infinite.
We shall divide the proof of \lemref{lem:converse} into three subsequences: $\{ p_{n} \}_{n \in \mathcal{I}_{1}}$, $\{ p_{n} \}_{n \in \mathcal{I}_{2}}$, and $\{ p_{n} \}_{n \in \mathcal{I}_{3}}$, since the limit superior is the supremum of the set of subsequential limits.

\subsubsection{When $\mathcal{I}_{1}$ is countably infinite}
\label{sect:I1}

Assuming that $\mathcal{I}_{1}$ is countably infinite, we now prove \lemref{lem:converse} for the subsequence $\{ p_{n} \}_{n \in \mathcal{I}_{1}}$, where note from \eqref{def:xn} that $\mathcal{I}_{1}$ is nonempty only if $p^{\ast} = p_{0} < \sigma$.
For simplicity, we shall write $\mathcal{I}_{1} = \{ n_{k} \}_{k = 1}^{\infty}$ in this subsubsection.
Firstly, it follows from the symbol-wise converse bound (cf.\ \cite[Proposition~6]{tomamichel_tan_2013} or \cite[Proposition~4.4]{tan_2014}) that
\begin{align}
\log M_{n}^{\ast}(\sigma, \varepsilon + \epsilon_{n})
\le
D_{\mathrm{s}}^{\varepsilon + \epsilon_{n} + \eta}(W_{n}^{n}(\cdot \mid \bvec{x}^{(n)}) \, \| \, Q^{(n)}) + \log \frac{ 1 }{ \eta }
\label{eq:symbol-wise_meta-converse}
\end{align}
for every $n \ge 1$.
Secondly, it follows by the sifting property of the information spectrum divergence from a convex combination $Q^{(n)}$ (cf.\ \cite[Lemma~3]{tomamichel_tan_2013} or \cite[Lemma~2.2]{tan_2014}) that
\begin{align}
D_{\mathrm{s}}^{\varepsilon + \epsilon_{n} + \eta}(W_{n}^{n}(\cdot \mid \bvec{x}^{(n)}) \, \| \, Q^{(n)})
\le
D_{\mathrm{s}}^{\varepsilon + \epsilon_{n} + \eta}(W_{n}^{n}(\cdot \mid \bvec{x}^{(n)}) \, \| \, (P_{[\kappa]}^{\ast}W_{n})^{n}) + \log 3
\label{eq:sifting_gamma}
\end{align}
where $\kappa$ is given in \eqref{def:kappa}.
Thirdly, it follows by Chebyshev's inequality (cf.\ \cite[Equation~(5)]{tomamichel_tan_2013} or \cite[Proposition~2.2]{tan_2014}) that
\begin{align}
D_{\mathrm{s}}^{\varepsilon + \epsilon_{n} + \eta}(W_{n}^{n}(\cdot \mid \bvec{x}^{(n)}) \, \| \, (P_{[\kappa]}^{\ast}W_{n})^{n})
\le
n \, D_{n} + \sqrt{ \frac{ n \, V_{n} }{ 1 - \varepsilon -\epsilon_{n} - \eta } } ,
\label{eq:chebyshev}
\end{align}
where $D_{n}$ and $V_{n}$ are given by
\begin{align}
D_{n}
& =
\frac{ 1 }{ n } \sum_{i = 1}^{n} D(W_{n}(\cdot \mid x_{i}^{(n)}) \, \| \, P_{[\kappa]}^{\ast}W_{n}) ,
\\
V_{n}
& =
\frac{ 1 }{ n } \sum_{i = 1}^{n} V(W_{n}(\cdot \mid x_{i}^{(n)}) \, \| \, P_{[\kappa]}^{\ast}W_{n}) .
\end{align}
As $P_{n}$ is the type of $\bvec{x}^{(n)} = (x_{1}^{(n)}, \dots, x_{n}^{(n)})$ (see \eqref{def:xn} and \eqref{def:P_xn}), note that
\begin{align}
D_{n}
& =
D(W_{n} \, \| \, P_{[\kappa]}^{\ast}W_{n} \mid P_{n}) ,
\label{eq:Dn_gamma_type} \\
V_{n}
& =
V(W_{n} \, \| \, P_{[\kappa]}^{\ast}W_{n} \mid P_{n}) .
\label{eq:Vn_gamma_type}
\end{align}
As shown in \appref{app:eq:asympt_Dn_gamma}, it holds that
\begin{align}
n \, D_{n}
=
T \, \tilde{C}( p_{n} ) + \mathrm{o}( 1 )
\label{eq:asympt_Dn_gamma}
\end{align}
as $n \to \infty$, where the mapping $\tilde{C} : [0, \sigma] \to [0, \infty)$ is defined by
\begin{align}
\tilde{C}( u )
& \coloneqq
A \, \bigg( (p^{\ast}+\kappa-u) + (1-u) \, s \log \frac{ s }{ p^{\ast}+\kappa+s } + u \, (1+s) \log \frac{ 1+s }{ p^{\ast}+\kappa+s } \bigg) .
\end{align}
After some algebra, it follows from the definition of $p_{0}$ in \eqref{def:p0} that
\begin{align}
\tilde{C}( u )
& =
A \, \bigg( (p^{\ast}+\kappa) - u \, \bigg( 1 + s \log \frac{ s }{ p^{\ast}+\kappa+s } - (1+s) \log \frac{ 1+s }{ p^{\ast}+\kappa+s } \bigg) + s \log \frac{ s }{ p^{\ast}+\kappa+s } \bigg)
\notag \\
& =
A \, \bigg( (p^{\ast}+\kappa) + u \log \frac{ (1+s)^{1+s} }{ s^{s} \, \mathrm{e} \, (p^{\ast}+\kappa+s) } + s \log \frac{ s }{ p^{\ast}+\kappa+s } \bigg)
\notag \\
& =
A \, \bigg( (p^{\ast}+\kappa) + u \log \frac{ p_{0}+s }{ p^{\ast}+\kappa+s } + s \log \frac{ s }{ p^{\ast}+\kappa+s } \bigg) ,
\label{eq:Ctilde}
\end{align}
where the last equality follows by the definition of $p_{0}$ in \eqref{def:p0}.
Since $p^{\ast} = p_{0} < \sigma$, equality \eqref{eq:Ctilde} implies that $\tilde{C} : [0, 1] \to \mathbb{R}$ is a strictly decreasing function on $[0, 1]$.
Thus, since $p^{\ast}+\kappa \le p_{n_{k}} \le \sigma$ for every $k \ge 1$, we have
\begin{align}
\limsup_{k \to \infty} \tilde{C}( p_{n_{k}} )
\le
\tilde{C}( p^{\ast}+\kappa )
<
C^{\ast} ,
\label{eq:tildeC_Cast}
\end{align}
where the second inequality is due to \cite[Equation~(2.11) in Part~I]{wyner_1988}.
On the other hand, as shown in \appref{app:eq:asympt_Vn_gamma}, it holds that
\begin{align}
\limsup_{n \to \infty} n \, V_{n}
\le
8 \, A \, T  \, (p^{\ast}+\kappa+s) .
\label{eq:asympt_Vn_gamma}
\end{align}
Combining \eqref{eq:symbol-wise_meta-converse}--\eqref{eq:chebyshev}, \eqref{eq:asympt_Dn_gamma}, \eqref{eq:tildeC_Cast}, and \eqref{eq:asympt_Vn_gamma}, we obtain
\begin{align}
\limsup_{k \to \infty} \log M_{n_{k}}^{\ast}(\sigma, \varepsilon + \epsilon_{n_{k}})
& <
T \, \tilde{C}( p^{\ast}+\kappa ) + 2 \, \sqrt{ \frac{ 2 \, A \, T \, (p^{\ast}+\kappa+s) }{ 1 - \varepsilon - \eta } } + \log \frac{ 1 }{ \eta } + \log 3
\notag \\
& =
T \, \tilde{C}( p^{\ast}+\kappa ) + \mathrm{O}( \sqrt{T} )
\notag \\
& \le
T \, C^{\ast} + \mathrm{o}( 1 )
\end{align}
as $T \to \infty$.
Therefore, \lemref{lem:converse} holds for the subsequence $\{ n_{k} \}_{k=1}^{\infty} = \mathcal{I}_{1}$.

\subsubsection{When $\mathcal{I}_{2}$ is countably infinite}
\label{sect:I2}

The proof can be done in the same way as \sectref{sect:I1} by replacing the input distribution $P_{[\kappa]}^{\ast}$ by $P_{[-\kappa]}^{\ast}$ in \eqref{eq:sifting_gamma}.
Then, replacing every occurrence of $p^{\ast} + \kappa$ by $p^{\ast} - \kappa$, we can verify that \lemref{lem:converse} holds for the subsequence $\{ n_{k} \}_{k=1}^{\infty} = \mathcal{I}_{2}$, provided that $\mathcal{I}_{2}$ is countably infinite.

\subsubsection{When $\mathcal{I}_{3}$ is countably infinite}
\label{sect:I3}

Assuming that $\mathcal{I}_{3}$ is countably infinite, we now prove \lemref{lem:converse} for the subsequence $\{ p_{n} \}_{n \in \mathcal{I}_{3}}$.
For brevity, we shall write $\mathcal{I}_{3} = \{ n_{k} \}_{k = 1}^{\infty}$ in this subsubsection.
Define the integer $m_{k}$ as
\begin{align}
m_{k}
\coloneqq
\min\bigg\{ m \in \mathbb{Z} \ \bigg| \ p_{n_{k}} \le p^{\ast} + \frac{ m }{ T } \bigg\}
\end{align}
for each $k \ge 1$.
Since  $p_{n_{k}} < p^{\ast} + \kappa$ for every $k \ge 1$ (see \eqref{eq:I3}), it follows that
\begin{align}
m_{k}
\le
\tilde{m}_{T}
\coloneqq
\min\bigg\{ m \in \mathbb{Z} \ \bigg| \ \kappa \le \frac{ m }{ T } \bigg\} .
\label{def:mk}
\end{align}
Moreover, since $\tilde{m}_{T}/T \to \kappa$ as $T \to \infty$, there exists a $T_{0} > 0$ satisfying
\begin{align}
\frac{ m_{k} }{ T }
\le
2 \, \kappa
<
1 - p^{\ast}
\label{eq:bound_mk}
\end{align}
for every $k \ge 1$ and $T \ge T_{0}$; henceforth, assume that $T \ge T_{0}$.
Denote by
\begin{align}
\tilde{P}_{k}
& \coloneqq
P_{[m_{k}/T]}^{\ast} ,
\\
\tilde{p}_{k}
& \coloneqq
\tilde{P}_{k}( 1 )
=
p^{\ast} + \frac{ m_{k} }{ T } ,
\label{def:p_tilde} \\
\tilde{r}_{k}
& \coloneqq
\tilde{P}_{k} W_{n_{k}}( 1 )
\label{def:rk_tilde}
\end{align}
for each $k \ge 1$.
By the definition of $\tilde{p}_{k}$ in \eqref{def:p_tilde}, it is clear that $\tilde{p}_{k}$ is bounded away from zero for all $k \ge 1$.
It follows by the sifting property of the information spectrum divergence from a convex combination $Q^{(n)}$ (cf.\ \cite[Lemma~3]{tomamichel_tan_2013} or \cite[Lemma~2.2]{tan_2014}) that
\begin{align}
D_{\mathrm{s}}^{\varepsilon + \epsilon_{n_{k}} + \eta}(W_{n_{k}}^{n_{k}}(\cdot \mid \bvec{x}^{(n_{k})}) \, \| \, Q^{(n_{k})})
& \le
D_{\mathrm{s}}^{\varepsilon + \epsilon_{n_{k}} + \eta}(W_{n_{k}}^{n_{k}}(\cdot \mid \bvec{x}^{(n_{k})}) \, \| \, (\tilde{P}_{k}W_{n_{k}})^{n_{k}}) + \log 3 + \log \left( 1 + \sqrt{ \frac{ \pi \, T }{ \gamma } } \right) + \frac{ \gamma \, m_{k}^{2} }{ T } .
\label{eq:sifting_from_a_convex_combination}
\end{align}
By the definition of $\tilde{m}_{T}$ in \eqref{def:mk}, it follows that
\begin{align}
m_{k}
\le
T \, |p_{n_{k}} - p^{\ast}| + 1 ;
\end{align}
that is, Inequality \eqref{eq:sifting_from_a_convex_combination} can be relaxed as
\begin{align}
D_{\mathrm{s}}^{\varepsilon + \epsilon_{n_{k}} + \eta}(W_{n_{k}}^{n_{k}}(\cdot \mid \bvec{x}^{(n_{k})}) \, \| \, Q^{(n_{k})})
& \le
D_{\mathrm{s}}^{\varepsilon + \epsilon_{n_{k}} + \eta}(W_{n_{k}}^{n_{k}}(\cdot \mid \bvec{x}^{(n_{k})}) \, \| \, (\tilde{P}_{k}W_{n_{k}})^{n_{k}}) + \log 3
\notag \\
& \qquad \qquad {}
+ \log \bigg( 1 + \sqrt{ \frac{ \pi \, T }{ \gamma } } \bigg) + \gamma \, \bigg( T \, (p_{n_{k}} - p^{\ast})^{2} + 2 \, |p_{n_{k}} - p^{\ast}| + \frac{ 1 }{ T } \bigg) .
\label{eq:sifting_mk_bound}
\end{align}
On the other hand, it follows by the Berry--Esseen theorem (cf.\ \cite[Lemma~5]{tomamichel_tan_2013} or \cite[Proposition~2.2]{tan_2014}) that
\begin{align}
D_{\mathrm{s}}^{\varepsilon + \epsilon_{n_{k}} + \eta}(W_{n_{k}}^{n_{k}}(\cdot \mid \bvec{x}^{(n_{k})}) \, \| \, (P^{\ast}W_{n_{k}})^{n_{k}})
& \le
n_{k} \, D_{k} + \sqrt{ n_{k} \, V_{k} } \, \Phi^{-1}\Bigg( \varepsilon + \epsilon_{n_{k}} + \eta + \frac{ 6 \, \Xi_{k} }{ \sqrt{ n_{k} \, V_{k}^{3} } } \Bigg) ,
\label{eq:Berry-Esseen_converse}
\end{align}
where $D_{k}$, $V_{k}$, and $\Xi_{k}$ are given by
\begin{align}
D_{k}
& =
\frac{ 1 }{ n_{k} } \sum_{i = 1}^{n_{k}} D(W_{n_{k}}(\cdot \mid x_{i}^{(n_{k})}) \, \| \, \tilde{P}_{k}W_{n_{k}}) ,
\\
V_{k}
& =
\frac{ 1 }{ n_{k} } \sum_{i = 1}^{n_{k}} V(W_{n_{k}}(\cdot \mid x_{i}^{(n_{k})}) \, \| \, \tilde{P}_{k}W_{n_{k}}) ,
\\
\Xi_{k}
& =
\frac{ 1 }{ n_{k} } \sum_{i = 1}^{n_{k}} \Xi(W_{n_{k}}(\cdot \mid x_{i}^{(n_{k})}) \, \| \, \tilde{P}_{k}W_{n_{k}}) .
\end{align}
As $P_{n}$ is the type of $\bvec{x}^{(n)} = (x_{1}^{(n)}, \dots, x_{n}^{(n)})$ (see \eqref{def:xn} and \eqref{def:P_xn}), it follows that
\begin{align}
D_{k}
& =
D(W_{n_{k}} \, \| \, \tilde{P}_{k}W_{n_{k}} \mid P_{n_{k}}) ,
\label{eq:Dn_type_ast} \\
V_{k}
& =
V(W_{n_{k}} \, \| \, \tilde{P}_{k}W_{n_{k}} \mid P_{n_{k}}) ,
\label{eq:Vn_type_ast} \\
\Xi_{k}
& =
\Xi(W_{n_{k}} \, \| \, \tilde{P}_{k}W_{n_{k}} \mid P_{n_{k}}) .
\label{eq:Tn_type_ast}
\end{align}
As shown in \appref{app:Dk_quadratic}, it holds that
\begin{align}
n_{k} \, D_{k}
\le
T \, C^{\ast} - T \, (p_{n_{k}} - p^{\ast})^{2} \, G_{1} + A + \mathrm{o}( 1 )
\label{eq:Dk_quadratic}
\end{align}
as $k \to \infty$, where the constant $G_{1} > 0$ (whose positivity is asserted and proved in \appref{app:Dk_quadratic}) is given by
\begin{align}
G_{1}
\coloneqq
A \, \bigg( \log \frac{ p_{0}+s }{ p^{\ast}+s } + \frac{ 1 }{ 2 \, (p^{\ast}+s) } \bigg( 1 - \frac{ \kappa }{ 3 \, (p^{\ast}+s) } \bigg) \bigg) .
\end{align}
Moreover, as shown in \appref{app:Vk_quadratic}, by exploiting the Lipschitz properties of the information variances, there exist two constants $\beta_{1}, \beta_{2} > 0$ satisfying
\begin{align}
\Big| \sqrt{ n_{k} \, V_{k} } - \sqrt{ T \, V^{\ast} } \Big|
\le
\frac{ \beta_{1} }{ \sqrt{T} } + \sqrt{T} \, |p_{n_{k}} - p^{\ast}| \, \beta_{2} + \mathrm{o}( 1 )
\label{eq:Vk_quadratic}
\end{align}
as $k \to \infty$.
Furthermore, as shown in \appref{app:Taylor_Phi_inverse}, assuming that $\eta = 1/\sqrt{T}$, there exist a constant $G_{2} > 0$ and a positive sequence $\delta_{k} = \mathrm{o}( 1 )$ (as $k \to \infty$) such that
\begin{align}
\Phi^{-1}\left( \varepsilon + \epsilon_{n_{k}} + \frac{ 1 }{ \sqrt{T} } + \frac{ 6 \, \Xi_{k} }{ \sqrt{ n_{k} \, V_{k}^{3} } } \right)
\le
\Phi^{-1}(\varepsilon + \epsilon_{n_{k}} + \delta_{k}) + \frac{ G_{2} }{ \sqrt{T} }
\label{eq:Taylor_Phi_inverse}
\end{align}
for sufficiently large $k$ and $T$.

Finally, we obtain
\begin{align}
\log M_{n_{k}}^{\ast}(\sigma, \varepsilon + \epsilon_{n_{k}})
& \overset{\mathclap{\text{(a)}}}{\le}
D_{\mathrm{s}}^{\varepsilon + \epsilon_{n_{k}} + (1/\sqrt{T})}(W_{n_{k}}^{n_{k}}(\cdot \mid \bvec{x}^{(n_{k})}) \, \| \, Q^{(n_{k})}) + \frac{ 1 }{ 2 } \log T
\notag \\
& \overset{\mathclap{\text{(b)}}}{\le}
D_{\mathrm{s}}^{\varepsilon + \epsilon_{n_{k}} + (1/\sqrt{T})}(W_{n_{k}}^{n_{k}}(\cdot \mid \bvec{x}^{(n_{k})}) \, \| \, (\tilde{P}_{k}W_{n_{k}})^{n_{k}}) + \log 3
\notag \\
& \qquad \qquad \qquad {}
+ \log \left( 1 + \sqrt{ \frac{ \pi \, T }{ \gamma } } \right) + \gamma \, \bigg( T \, (p_{n_{k}} - p^{\ast})^{2} + 2 + \frac{ 1 }{ T } \bigg)  + \frac{ 1 }{ 2 } \log T
\notag \\
& \quad \overset{\mathclap{\text{(c)}}}{\le}
n_{k} \, D_{k} + \sqrt{ n_{k} \, V_{k} } \, \Phi^{-1}\left( \varepsilon + \epsilon_{n_{k}} + \frac{ 1 }{ \sqrt{T} } + \frac{ 6 \, \Xi_{k} }{ \sqrt{ n_{k} \, V_{k}^{3} } } \right) + \log 3
\notag \\
& \qquad \qquad \qquad {}
+ \log \left( 1 + \sqrt{ \frac{ \pi \, T }{ \gamma } } \right) + \gamma \, \bigg( T \, (p_{n_{k}} - p^{\ast})^{2} + 2 + \frac{ 1 }{ T } \bigg)  + \frac{ 1 }{ 2 } \log T
\notag \\
& \overset{\mathclap{\text{(d)}}}{\le}
n_{k} \, D_{k} + \sqrt{ n_{k} \, V_{k} } \, \bigg( \Phi^{-1}(\varepsilon + \epsilon_{n_{k}} + \delta_{k}) + \frac{ G_{2} }{ \sqrt{T} } \bigg) + \log 3
\notag \\
& \qquad \qquad \qquad {}
+ \log \left( 1 + \sqrt{ \frac{ \pi \, T }{ \gamma } } \right) + \gamma \, \bigg( T \, (p_{n_{k}} - p^{\ast})^{2} + 2  + \frac{ 1 }{ T } \bigg)  + \frac{ 1 }{ 2 } \log T
\notag \\
& \overset{\mathclap{\text{(e)}}}{\le}
\Big( T \, C^{\ast} - T \, (p_{n_{k}} - p^{\ast})^{2} \, G_{1} + A \Big) + \sqrt{ n_{k} \, V_{k} } \, \bigg( \Phi^{-1}(\varepsilon + \epsilon_{n_{k}} + \delta_{k}) + \frac{ G_{2} }{ \sqrt{T} } \bigg) + \log 3
\notag \\
& \qquad \qquad \qquad {}
+ \log \left( 1 + \sqrt{ \frac{ \pi \, T }{ \gamma } } \right) + \gamma \, \bigg( T \, (p_{n_{k}} - p^{\ast})^{2} + 2 + \frac{ 1 }{ T } \bigg) + \frac{ 1 }{ 2 } \log T
\notag \\
& \overset{\mathclap{\text{(f)}}}{\le}
T \, C^{\ast} - T \, (p_{n_{k}} - p^{\ast})^{2} \, G_{1} + A + \sqrt{ T \, V^{\ast} } \, \Phi^{-1}(\varepsilon + \epsilon_{n_{k}} + \delta_{k}) + G_{2} \, \bigg( \sqrt{V^{\ast}} + \frac{ \beta_{1} }{ T } + \beta_{2} \bigg)
\notag \\
& \quad \qquad {}
+ \bigg( \frac{ \beta_{1} }{ T } + \sqrt{T} \, |p_{n_{k}} - p^{\ast}| \, \beta_{2} \bigg) \, |\Phi^{-1}(\varepsilon + \epsilon_{n_{k}} + \delta_{k})|  + \log 3
\notag \\
& \qquad \qquad \qquad {}
+ \log \left( 1 + \sqrt{ \frac{ \pi \, T }{ \gamma } } \right) + \gamma \, \bigg( T \, (p_{n_{k}} - p^{\ast})^{2} + 2 + \frac{ 1 }{ T } \bigg)  + \frac{ 1 }{ 2 } \log T + \mathrm{o}( 1 )
\notag \\
& \overset{\mathclap{\text{(g)}}}{\le}
T \, C^{\ast} + \sqrt{ T \, V^{\ast} } \, \Phi^{-1}(\varepsilon + \epsilon_{n_{k}} + \delta_{k}) + \frac{ 1 }{ 2 } \log T + \log \left( 1 + \sqrt{ \frac{ \pi \, T }{ \gamma } } \right)
\notag \\
& \qquad {}
+ f\Big( \sqrt{T} \, |p_{n_{k}} - p^{\ast}| \Big) + \frac{ \beta_{1} \, (|\Phi^{-1}(\varepsilon + \epsilon_{n_{k}} + \delta_{k})| + G_{2}) + \gamma }{ T } + \mathrm{o}( 1 )
\label{eq:before_quadratic}
\end{align}
as $k \to \infty$ and for sufficiently large $T$, where
\begin{itemize}
\item
(a) follows from \eqref{eq:symbol-wise_meta-converse},
\item
(b) follows from \eqref{eq:sifting_mk_bound},
\item
(c) follows from \eqref{eq:Berry-Esseen_converse},
\item
(d) follows from \eqref{eq:Taylor_Phi_inverse},
\item
(e) follows from \eqref{eq:Dk_quadratic},
\item
(f) follows from \eqref{eq:Vk_quadratic}, and
\item
(g) follows by defining the function $f : \mathbb{R} \to \mathbb{R}$ as
\begin{align}
f( u )
& \coloneqq
u^{2} \, (\gamma - G_{1}) + u \, \beta_{2} \, |\Phi^{-1}(\varepsilon + \epsilon_{n_{k}} + \delta_{k})| + G_{2} \, \Big( \sqrt{V^{\ast}} + \beta_{2} \Big) + A + \log 3 + 2 \, \gamma .
\label{def:quadratic}
\end{align}
\end{itemize}
By choosing the constant $\gamma > 0$ in \eqref{eq:artifical} sufficiently small so that
\begin{align}
\gamma < G_{1} ,
\label{eq:choice_of_gamma}
\end{align}
it follows by the maximization of the quadratic function \eqref{def:quadratic}, which has a negative leading coefficient, that
\begin{align}
f\Big( \sqrt{T} \, |p_{n_{k}} - p^{\ast}| \Big)
& \le
\frac{ \beta_{2}^{2} \, \Phi^{-1}(\varepsilon + \epsilon_{n_{k}} + \delta_{k})^{2} }{ 4 \, (G_{1} - \gamma) } + G_{2} \, \Big( \sqrt{V^{\ast}} + \beta_{2} \Big) + A + \log 3 + 2 \, \gamma
\label{eq:quadratic}
\end{align}
for sufficiently large $k$ and $T$.
Combining \eqref{eq:before_quadratic} and \eqref{eq:quadratic}, we have
\begin{align}
\limsup_{k \to \infty} \log M_{n_{k}}^{\ast}(\sigma, \varepsilon + \epsilon_{n_{k}})
& \le
T \, C^{\ast} + \sqrt{ T \, V^{\ast} } \, \Phi^{-1}( \varepsilon ) + \log T + \mathrm{O}( 1 )
\end{align}
as $T \to \infty$, proving \lemref{lem:converse} for the subsequence $\{ n_{k} \}_{k=1}^{\infty} =\mathcal{I}_{3}$.
This completes the proof of \lemref{lem:converse}
\end{IEEEproof}

\subsection{Proof of Achievability Part of \thref{th:2nd-order}}
\label{sect:direct}

As shown in \sectref{sect:discretization}, a channel coding problem for the Poisson channel can be reduced to that of a certain discrete memoryless channel under an additive cost constraint.
Now, define a channel code under an average cost-constraint as follows:

\begin{definition}
\label{def:avg_cost}
Given a cost function $\chi : \mathcal{X} \to \mathbb{R}$, a pair of encoder $f : \{ 1, \dots, M \} \to \mathcal{X}$ and decoder $g : \mathcal{Y} \to \{ 1, \dots, M \}$ is called an \emph{$(M, \varepsilon, \beta)_{\mathrm{avg}}$-code for a channel $W : \mathcal{X} \to \mathcal{Y}$} if
\begin{align}
\frac{ 1 }{ M } \sum_{m = 1}^{M} W(g^{-1}(m) \mid f(m))
& \ge
1 - \varepsilon ,
\end{align}
and
\begin{align}
\max_{m \in \{ 1, \dots, M \}} \chi \circ f( m )
& \le
\beta .
\end{align}
\end{definition}

The following lemma is a straightforward generalization of the random coding union bound \cite[Theorem~16]{polyanskiy_poor_verdu_2010}.

\begin{lemma}
\label{lem:RCU_cost}
Let $M$ be a positive integer and $\beta$ a real.
For any distribution $P$ on $\mathcal{X}$ and any channel $W : \mathcal{X} \to \mathcal{Y}$, there exists an $(M, \varepsilon, \beta)_{\mathrm{avg}}$-code satisfying
\begin{align}
\varepsilon
& \le
\mathbb{P}\{ \chi(X) > \beta \} + \mathbb{E}\bigg[ \min\bigg\{ 1, (M-1) \, \mathbb{P}\bigg\{ \log \frac{ W(Y \mid \bar{X}) }{ PW( Y ) } \ge \log \frac{ W(Y \mid X) }{ PW( Y ) } \ \bigg| \ X, Y \bigg\} \bigg\} \bigg] ,
\label{eq:RCU_cost}
\end{align}
where the r.v.'s $X$, $Y$, and $\bar{X}$ are defined so that $\mathbb{P} \circ (X, Y)^{-1} = P \times W$, $\mathbb{P} \circ \bar{X}^{-1} = P$, and $(X, Y) \Perp \bar{X}$.
\end{lemma}

\begin{IEEEproof}[Proof of \lemref{lem:RCU_cost}]
See \appref{app:RCU_cost}.
\end{IEEEproof}

To ensure that $\Delta_{n} = \mathrm{o}( 1 )$ as $T \to \infty$, assume throughout the achievability proof that%
\footnote{As a matter of fact, it suffices to assume in the achievability proof that $n = \omega(T)$ as $T \to \infty$.
We assume, however, here that \eqref{def:n_omegaT} holds to give an explicit dependence between $n$ and $T$.
This dependence affects some constants that appear later in the proof.}
\begin{align}
n
=
\lceil T^{2} \rceil ,
\label{def:n_omegaT}
\end{align}
where $\Delta_{n}$ is defined in \eqref{def:Delta_n}, and $\lceil u \rceil \coloneqq \min\{ z \in \mathbb{Z} \mid z \ge u \}$ stands for the ceiling function.
Recall that the $u$-shifted distribution $P_{[u]}^{\ast}$ from the CAID $P^{\ast}$ is defined in \eqref{def:u-shift}.
For each $n \ge 1$, define the Bernoulli distribution $P_{n}$ by%
\footnote{The choice of parameter $u_{n} = -p^{\ast} \, n^{-1/4}$ is inspired by the delta-convention \cite[Convention~2.11]{csiszar_korner_2011}.}
\begin{align}
P_{n}
\coloneqq
P_{[u_{n}]}^{\ast}
\qquad (\mathrm{with} \ u_{n} = -p^{\ast} \, n^{-1/4}) .
\label{def:delta-convention}
\end{align}
It is clear that $P_{n}$ converges to $P^{\ast}$ in the variational distance topology as $T \to \infty$.
Denote by
\begin{align}
p_{n}
& \coloneqq
P_{n}( 1 ) ,
\\
r_{n}
& \coloneqq
P_{n}W_{n}( 1 )
\label{def:output_rn_direct}
\end{align}
the Bernoulli parameters of $P_{n}$ and $P_{n}W_{n}$, respectively.
Now, we have the following lemma.

\begin{lemma}
\label{lem:unconditional_divergences}
For $n$ satisfying \eqref{def:n_omegaT}, the following asymptotic estimates hold:
\begin{align}
n \, I(P_{n}, W_{n})
& =
T \, C^{\ast} + \mathrm{o}( 1 ) ,
\label{eq:asympt_I_direct} \\
n \, \tilde{V}(P_{n}, W_{n})
& =
T \, V^{\ast} + \mathrm{o}( 1 ) ,
\label{eq:asympt_V_direct} \\
n \, \tilde{\Xi}(P_{n}, W_{n})
& =
T \, \Xi^{\ast} + \mathrm{o}( 1 )
\label{eq:asympt_Xi_direct}
\end{align}
as $T \to \infty$, where $C^{\ast}$ and $V^{\ast}$ are defined in \eqref{def:capacity_Poisson} and \eqref{def:dispersion_Poisson}, respectively, and $\Xi^{\ast}$ is defined as
\begin{align}
\Xi^{\ast}
& \coloneqq
A \, \bigg( p^{\ast} \, (1+s) \log^{3} \frac{ 1+s }{ p^{\ast}+s } - (1 - p^{\ast}) \, s \log^{3} \frac{ s }{ p^{\ast}+s } \bigg) .
\label{def:third-moment_Poisson}
\end{align}
\end{lemma}

\begin{IEEEproof}[Proof of \lemref{lem:unconditional_divergences}]
See \appref{app:unconditional_divergences}.
\end{IEEEproof}

For each $n \ge 1$, denote by
\begin{align}
\color{blue}
(X_{n,1}, Y_{n, 1}), (X_{n, 2}, Y_{n, 2}), \dots, (X_{n, n}, Y_{n, n})
\end{align}
i.i.d.\ pairs of r.v.'s with generic distribution $P_{n} \times W_{n}$.
For short, we write the random vectors
\begin{align}
X^{n}
& =
(X_{n, 1}, X_{n, 2}, \dots, X_{n, n}) ,
\\
Y^{n}
& =
(Y_{n, 1}, Y_{n, 2}, \dots, Y_{n, n}) .
\end{align}

\begin{lemma}
\label{lem:delta-convention}
It holds that
\begin{align}
\mathbb{P}\{ N(1 \mid X^{n}) > n \, \sigma \}
\le
\mathrm{e}^{- 2 (p^{\ast})^{2} T}
\end{align}
for sufficiently large $T$, where $N(a \mid \bvec{z})$ denotes the number of occurrences of an element $a \in \mathcal{A}$ in the sequence $\bvec{z} = (z_{1}, \dots, z_{m}) \in \mathcal{A}^{m}$, i.e.,
\begin{align}
N(a \mid \bvec{z})
\coloneqq
\Big|\Big\{ 1 \le i \le m \ \Big| \ z_{i} = a \Big\}\Big| .
\end{align}
\end{lemma}

\begin{IEEEproof}[Proof of \lemref{lem:delta-convention}]
See \appref{app:delta-convention}.
\end{IEEEproof}

Given a constant $0 < \kappa < 1$, define the event
\begin{align}
\mathcal{E}_{T}
\coloneqq
\Big\{ N(1 \mid Y^{n}) > \mathbb{E}[ N(1 \mid Y^{n}) ] - \kappa \, (p^{\ast}+s) \, A \, T \Big\}
\label{def:event_n}
\end{align}
for each $T > 0$, where note that
\begin{align}
\mathbb{E}[ N(1 \mid Y^{n}) ]
=
\mathbb{E}\bigg[ \sum_{i = 1}^{n} Y_{n, i} \bigg]
=
n \, r_{n} ,
\label{eq:expectation_1Y}
\end{align}
because $Y_{n, 1}, \dots, Y_{n, n}$ are i.i.d.\ Bernoulli r.v.'s with the same parameter $r_{n}$.
The following lemma asserts that \eqref{def:event_n} is a high probability set asymptotically in the sense that the probability of its complement decays exponentially fast in $T$.

\begin{lemma}
\label{lem:eventEn_complement}
There exists a positive constant $K_{0} = K_{0}(\kappa, \lambda_{0}, A, \sigma)$ satisfying
\begin{align}
\mathbb{P}( \mathcal{E}_{T}^{\complement} )
\le
\mathrm{e}^{ - K_{0}  T }
\end{align}
for sufficiently large $T$, where $\mathcal{S}^{\complement}$ stands for the complement of the set $\mathcal{S}$.
\end{lemma}

\begin{IEEEproof}[Proof of \lemref{lem:eventEn_complement}]
See \appref{app:eventEn_complement}.
\end{IEEEproof}

For each $n \ge 1$ and $\bvec{x}, \bvec{y} \in \{ 0, 1 \}^{n}$, define the information density
\begin{align}
\iota_{n}(\bvec{x} \wedge \bvec{y})
\coloneqq
\log \frac{ W_{n}^{n}(\bvec{y} \mid \bvec{x}) }{ (P_{n}W_{n})^{n}( \bvec{y} ) } .
\end{align}
Then, the following two lemmas, which are analogous to \cite[Lemma~47]{polyanskiy_poor_verdu_2010}, hold.

\begin{lemma}
\label{lem:tight_3rd-order_term}
There exists a positive constant $K_{1} = K_{1}(\kappa, \lambda_{0}, A, \sigma)$ such that
\begin{align}
\bvec{1}_{\mathcal{E}_{T}} \, \mathbb{E}\Big[ \mathrm{e}^{- \iota_{n}(X^{n} \wedge Y^{n})} \, \bvec{1}\{ \iota_{n}(X^{n} \wedge Y^{n}) \ge \gamma \} \ \Big| \ Y^{n} \Big]
\le
\frac{ K_{1} \, \mathrm{e}^{- \gamma} }{ \sqrt{ T } }
\end{align}
almost surely for every real $\gamma$ and for sufficiently large $T$, where $\bvec{1}_{\mathcal{E}}$ stands for the indicator function of the event $\mathcal{E}$.
\end{lemma}

\begin{IEEEproof}[Proof of \lemref{lem:tight_3rd-order_term}]
See \appref{app:tight_3rd-order_term}.
\end{IEEEproof}

\begin{lemma}
\label{lem:final_sqrt}
There exists a positive constant $K_{2} = K_{2}(\lambda_{0}, A, \sigma)$ satisfying
\begin{align}
\mathbb{E} \Big[ \mathrm{e}^{-\iota_{n}(X^{n} \wedge Y^{n})} \, \bvec{1}\{ \iota_{n}(X^{n} \wedge Y^{n}) > \gamma \} \Big]
\le
\frac{ K_{2} \, \mathrm{e}^{-\gamma} }{ \sqrt{T} }
\end{align}
for every real $\gamma$ and for sufficiently large $T$.
\end{lemma}

\begin{IEEEproof}[Proof of \lemref{lem:final_sqrt}]
See \appref{app:final_sqrt}.
\end{IEEEproof}

The following lemma is a final tool to prove the achievability part of \thref{th:2nd-order}.

\begin{lemma}
\label{lem:Berry-Esseen_unconditional}
There exists a positive constant $K_{3} = K_{3}(\lambda_{0}, A, \sigma)$ satisfying
\begin{align}
\mathbb{P}\{ \iota_{n}(X^{n} \wedge Y^{n}) \le \gamma \}
\le
\Phi\left( \frac{ \gamma - n \, I(P_{n}, W_{n}) }{ \sqrt{ n \, \tilde{V}(P_{n}, W_{n}) } } \right) + \frac{ K_{3} }{ \sqrt{T} }
\end{align}
for every real $\gamma$ and for sufficiently large $T$.
\end{lemma}

\begin{IEEEproof}[Proof of \lemref{lem:Berry-Esseen_unconditional}]
See \appref{app:Berry-Esseen_unconditional}.
\end{IEEEproof}

Now, for each $n \ge 1$, we define the following numbers:
\begin{align}
M_{n}
& \coloneqq
\big\lceil \exp( S_{n} + G_{n} ) \big\rceil ,
\label{def:Mn} \\
S_{n}
& \coloneqq
n \, I(P_{n}, W_{n}) + \sqrt{ n \, \tilde{V}(P_{n}, W_{n}) } \, \Phi^{-1}( \varepsilon_{n} ) ,
\label{def:Sn} \\
G_{n}
& \coloneqq
\frac{ 1 }{ 2 } \log T - \log K_{1} ,
\label{def:Gn} \\
\varepsilon_{n}
& \coloneqq
\varepsilon - \frac{ K_{2} + K_{3} }{ \sqrt{T} } - \mathrm{e}^{- 2 (p^{\ast})^{2} T} - \mathrm{e}^{- K_{0} T} ,
\label{def:epsn}
\end{align}
where the constants $K_{0}$, $K_{1}$, $K_{2}$, and $K_{3}$ are given in Lemmas~\ref{lem:eventEn_complement}--\ref{lem:Berry-Esseen_unconditional}, respectively.
Then, it follows from \lemref{lem:unconditional_divergences} that
\begin{align}
\log M_{n}
& \ge
n \, I(P_{n}, W_{n}) + \sqrt{ n \, \tilde{V}(P_{n}, W_{n}) } \, \Phi^{-1}\bigg( \varepsilon - \frac{ K_{2} + K_{3} }{ \sqrt{T} } - \mathrm{e}^{- 2 (p^{\ast})^{2} T} - \mathrm{e}^{- K_{0} T} \bigg) + \frac{ 1 }{ 2 } \log T - \log K_{1}
\notag \\
& \overset{\mathclap{\text{(a)}}}{=}
(T \, C^{\ast}+\mathrm{o}(1)) + \sqrt{ T \, V^{\ast} + \mathrm{o}( 1 ) } \, \bigg( \Phi^{-1}( \varepsilon ) + \mathrm{O}\bigg( \frac{ 1 }{ \sqrt{T} } \bigg) \bigg) + \frac{ 1 }{ 2 } \log T + \mathrm{O}( 1 )
\notag \\
& =
T \, C^{\ast} + \sqrt{ T \, V^{\ast} } \, \Phi^{-1}( \varepsilon ) + \frac{ 1 }{ 2 } \log T + \mathrm{O}( 1 )
\label{eq:asympt_Mn_RCU}
\end{align}
as $T \to \infty$, where in (a), we have also applied a Taylor series expansion of $\Phi^{-1}( \cdot )$ around $\varepsilon$.

Let $\bar{X}^{n}$ be a r.v.\ satisfying $\mathbb{P} \circ (\bar{X}^{n})^{-1} = P_{n}^{n}$ and $\bar{X}^{n} \Perp (X^{n}, Y^{n})$.
It follows from \lemref{lem:RCU_cost} that there exists an $(n, M_{n}, \sigma, \varepsilon^{\prime})_{\mathrm{avg}}$-code for the discretized channel $W_{n}^{n}$ such that 
\begin{align}
\varepsilon^{\prime}
& \le
\mathbb{E} \big[ \min\big\{ 1, (M_{n} - 1) \, \mathbb{P}\{ \iota_{n}(\bar{X}^{n} \wedge Y^{n}) \ge \iota_{n}(X^{n} \wedge Y^{n}) \mid X^{n}, Y^{n} \} \big\} \big] + \mathbb{P}\bigg\{ \frac{ 1 }{ n } \sum_{i=1}^{n} X_{n, i} > \sigma \bigg\}
\notag \\
& \overset{\mathclap{\text{(a)}}}{\le}
\mathbb{E} \big[ \min\big\{ 1, (M_{n} - 1) \, \mathbb{P}\{ \iota_{n}(\bar{X}^{n} \wedge Y^{n}) \ge \iota_{n}(X^{n} \wedge Y^{n}) \mid X^{n}, Y^{n} \} \big\} \big] + \mathrm{e}^{- 2 (p^{\ast})^{2} T}
\notag \\
& =
\mathbb{E} \Bigg[ \min\bigg\{ 1, (M_{n} - 1) \sum_{\bar{\bvec{x}} \in \{ 0, 1 \}^{n}} P_{\bar{X}^{n}}( \bar{\bvec{x}} ) \, \bvec{1}\{ \iota_{n}(\bar{\bvec{x}} \wedge Y^{n}) \ge \iota_{n}(X^{n} \wedge Y^{n}) \} \bigg\} \Bigg] + \mathrm{e}^{- 2 (p^{\ast})^{2} T}
\notag \\
& =
\mathbb{E} \Bigg[ \min\bigg\{ 1, (M_{n} - 1) \sum_{\bar{\bvec{x}} \in \{ 0, 1 \}^{n}} \mathbb{P}\{ X^{n} = \bar{\bvec{x}} \mid Y^{n} \} \frac{ (PW_{n})^{n}( Y^{n} ) }{ W_{n}^{n}(Y^{n} \mid \bar{\bvec{x}}) } \, \bvec{1}\{ \iota_{n}(\bar{\bvec{x}} \wedge Y^{n}) \ge \iota_{n}(X^{n} \wedge Y^{n}) \} \bigg\} \Bigg] + \mathrm{e}^{- 2 (p^{\ast})^{2} T}
\notag \\
& =
\mathbb{E} \Bigg[ \min\bigg\{ 1, (M_{n} - 1) \sum_{\bar{\bvec{x}} \in \{ 0, 1 \}^{n}} \mathbb{P}\{ X^{n} = \bar{\bvec{x}} \mid Y^{n} \} \exp( - \iota_{n}(\bar{\bvec{x}} \wedge Y^{n}) ) \, \bvec{1}\{ \iota_{n}(\bar{\bvec{x}} \wedge Y^{n}) \ge \iota_{n}(X^{n} \wedge Y^{n}) \} \bigg\} \Bigg] + \mathrm{e}^{- 2 (p^{\ast})^{2} T}
\notag \\
& =
\mathbb{E} \Bigg[ \bvec{1}_{\mathcal{E}_{T}} \, \min\bigg\{ 1, (M_{n} - 1) \sum_{\bar{\bvec{x}} \in \{ 0, 1 \}^{n}} \mathbb{P}\{ X^{n} = \bar{\bvec{x}} \mid Y^{n} \} \exp( - \iota_{n}(\bar{\bvec{x}} \wedge Y^{n}) ) \, \bvec{1}\{ \iota_{n}(\bar{\bvec{x}} \wedge Y^{n}) \ge \iota_{n}(X^{n} \wedge Y^{n}) \} \bigg\} \Bigg]
\notag \\
& \qquad
{} + \mathbb{E} \Bigg[ \bvec{1}_{\mathcal{E}_{T}^{\complement}} \, \min\bigg\{ 1, (M_{n} - 1) \sum_{\bar{\bvec{x}} \in \{ 0, 1 \}^{n}} \mathbb{P}\{ X^{n} = \bar{\bvec{x}} \mid Y^{n} \} \exp( - \iota_{n}(\bar{\bvec{x}} \wedge Y^{n}) ) \, \bvec{1}\{ \iota_{n}(\bar{\bvec{x}} \wedge Y^{n}) \ge \iota_{n}(X^{n} \wedge Y^{n}) \} \bigg\} \Bigg] + \mathrm{e}^{- 2 (p^{\ast})^{2} T}
\notag \\
& \le
\mathbb{E} \Bigg[ \bvec{1}_{\mathcal{E}_{T}} \, \min\bigg\{ 1, (M_{n} - 1) \sum_{\bar{\bvec{x}} \in \{ 0, 1 \}^{n}} \mathbb{P}\{ X^{n} = \bar{\bvec{x}} \mid Y^{n} \} \exp( - \iota_{n}(\bar{\bvec{x}} \wedge Y^{n}) ) \, \bvec{1}\{ \iota_{n}(\bar{\bvec{x}} \wedge Y^{n}) \ge \iota_{n}(X^{n} \wedge Y^{n}) \} \bigg\} \Bigg] + \mathrm{e}^{- 2 (p^{\ast})^{2} T} + \mathbb{P}( \mathcal{E}_{T}^{\complement} )
\notag \\
& \overset{\mathclap{\text{(b)}}}{\le}
\mathbb{E} \Bigg[ \bvec{1}_{\mathcal{E}_{T}} \, \min\bigg\{ 1, (M_{n} - 1) \sum_{\bar{\bvec{x}} \in \{ 0, 1 \}^{n}} \mathbb{P}\{ X^{n} = \bar{\bvec{x}} \mid Y^{n} \} \exp( - \iota_{n}(\bar{\bvec{x}} \wedge Y^{n}) ) \, \bvec{1}\{ \iota_{n}(\bar{\bvec{x}} \wedge Y^{n}) \ge \iota_{n}(X^{n} \wedge Y^{n}) \} \bigg\} \Bigg] + \mathrm{e}^{- 2 (p^{\ast})^{2} T} + \mathrm{e}^{- K_{0} T}
\notag \\
& \overset{\mathclap{\text{(c)}}}{\le}
\mathbb{E} \Bigg[ \min\bigg\{ 1, \frac{ K_{1} \, (M_{n} - 1) \, \exp( - \iota_{n}(X^{n} \wedge Y^{n}) ) }{ \sqrt{ T } } \bigg\} \Bigg] + \mathrm{e}^{- 2 (p^{\ast})^{2} T} + \mathrm{e}^{- K_{0} T}
\notag \\
& =
\mathbb{E} \Bigg[ \bvec{1}\bigg\{ \iota_{n}(X^{n} \wedge Y^{n}) \le \log \frac{ K_{1} \, (M_{n} - 1) }{ \sqrt{T} } \bigg\} \, \min\bigg\{ 1, \frac{ K_{1} \, (M_{n} - 1) \, \exp( - \iota_{n}(X^{n} \wedge Y^{n}) ) }{ \sqrt{ T } } \bigg\} \Bigg]
\notag \\
& \qquad
{} + \mathbb{E} \Bigg[ \bvec{1}\bigg\{ \iota_{n}(X^{n} \wedge Y^{n}) > \log \frac{ K_{1} \, (M_{n} - 1) }{ \sqrt{T} } \bigg\} \, \min\bigg\{ 1, \frac{ K_{1} \, (M_{n} - 1) \, \exp( - \iota_{n}(X^{n} \wedge Y^{n}) ) }{ \sqrt{ T } } \bigg\} \Bigg] + \mathrm{e}^{- 2 (p^{\ast})^{2} T} + \mathrm{e}^{- K_{0} T}
\notag \\
& \le
\mathbb{P}\bigg\{ \iota_{n}(X^{n} \wedge Y^{n}) \le \log \frac{ K_{1} \, (M_{n} - 1) }{ \sqrt{T} } \bigg\} + \frac{ K_{1} \, (M_{n}-1) }{ \sqrt{T} } \, \mathbb{E} \Bigg[ \bvec{1}\bigg\{ \iota_{n}(X^{n} \wedge Y^{n}) > \log \frac{ K_{1} \, (M_{n} - 1) }{ \sqrt{T} } \bigg\} \, \exp( - \iota_{n}(X^{n} \wedge Y^{n}) ) \Bigg] \nonumber\\*
&\qquad\qquad + \mathrm{e}^{- 2 (p^{\ast})^{2} T} + \mathrm{e}^{- K_{0} T}
\notag \\
& \overset{\mathclap{\text{(d)}}}{\le}
\mathbb{P}\bigg\{ \iota_{n}(X^{n} \wedge Y^{n}) \le \log \frac{ K_{1} \, (M_{n} - 1) }{ \sqrt{T} } \bigg\} + \frac{ K_{2} }{ \sqrt{T} } + \mathrm{e}^{- 2 (p^{\ast})^{2} T} + \mathrm{e}^{- K_{0} T}
\notag \\
& \overset{\mathclap{\text{(e)}}}{\le}
\Phi\left( \frac{ \log(M_{n}-1) + \log K_{1} - (1/2) \log T - n \, I(P_{n}, W_{n}) }{ \sqrt{ n \, \tilde{V}(P_{n}, W_{n}) } } \right) + \frac{ K_{2} + K_{3} }{ \sqrt{T} } + \mathrm{e}^{- 2 (p^{\ast})^{2} T} + \mathrm{e}^{- K_{0} T}
\notag \\
& \overset{\mathclap{\text{(f)}}}{\le}
\Phi\left( \frac{ S_{n} + G_{n} + \log K_{1} - (1/2) \log T - n \, I(P_{n}, W_{n}) }{ \sqrt{ n \, \tilde{V}(P_{n}, W_{n}) } } \right) + \frac{ K_{2} + K_{3} }{ \sqrt{T} } + \mathrm{e}^{- 2 (p^{\ast})^{2} T} + \mathrm{e}^{- K_{0} T}
\notag \\
& \overset{\mathclap{\text{(g)}}}{=}
\varepsilon
\label{eq:long-eqn}
\end{align}
for sufficiently large $T$, where
\begin{itemize}
\item
(a) follows from \lemref{lem:delta-convention},
\item
(b) follows from \lemref{lem:eventEn_complement},
\item
(c) follows from \lemref{lem:tight_3rd-order_term},
\item
(d) follows from \lemref{lem:final_sqrt},
\item
(e) follows from \lemref{lem:Berry-Esseen_unconditional},
\item
(f) follows by the definition of $M_{n}$ in \eqref{def:Mn}, and
\item
(g) follows by the definitions of $S_{n}$, $G_{n}$, and $\varepsilon_{n}$ in \eqref{def:Sn}--\eqref{def:epsn}, respectively.
\end{itemize}
Therefore, it follows from \eqref{eq:comparison} and \eqref{eq:asympt_Mn_RCU} that the achievability bound of \thref{th:2nd-order} is satisfied, completing the proof.
\hfill\IEEEQEDhere

\section{Concluding Remarks}
\label{sect:conclusion}

We have derived the optimal second-order coding rate for the continuous-time Poisson channel.
We have also obtained bounds on the third-order coding rate.
This is the first instance of a second-order asymptotic result for continuous-time communication models in information theory.
While the high-level proof ideas of \thref{th:2nd-order} are based on Wyner's discretization argument \cite{wyner_1988} and standard techniques in second-order asymptotics \cite{strassen_1962, hayashi_2009, polyanskiy_poor_verdu_2010, polyanskiy_thesis, tomamichel_tan_2013, tan_2014}, several novel finite blocklength techniques have to be introduced in both the converse and achievability arguments in view of the continuous-time nature of the Poisson channel.
In the following two subsections, we summarize these technical contributions, partitioning our discussion into the converse and achievability parts of \thref{th:2nd-order}.

\subsection{Technical Contributions in Proving Converse Part of \thref{th:2nd-order}}
\label{sect:comment_converse}

In the proof of \lemref{lem:converse}, we constructed a somewhat artificial output distribution $Q^{(n)}$ in \eqref{eq:artifical} to be substituted into the $\epsilon$-information spectrum divergence in \eqref{def:xn}.
This construction of our choice of $Q^{(n)}$ is partly inspired by Tomamichel and Tan's choice of the output distribution \cite[Equation~(6)]{tomamichel_tan_2013}, which was used to derive a tight third-order converse term for the fundamental limit of DMCs having positive channel dispersions \cite[Proposition~8]{tomamichel_tan_2013} (see also \cite[Section~4.2.3]{tan_2014}).
Tomamichel and Tan's choice is a hybrid between Hayashi's choice of the output distribution used in \cite[Section~X-A]{hayashi_2009} and another artificial output distribution based on the construction of an appropriate $\epsilon$-net in the output probability simplex.
Tomamichel and Tan's and Hayashi's choices cannot, however, be used to prove \lemref{lem:converse} even up to the second-order converse term $\sqrt{T \, V^{\ast}} \, \Phi^{-1}( \varepsilon )$ in \eqref{eq:converse}.
This is because both choices require the application of the type counting lemma \cite[Lemma~2.2]{csiszar_korner_2011} (see also \cite[Problem~2.1]{csiszar_korner_2011}) in the approximation arguments.
In contrast, due to the continuous nature of the Poisson channel and Wyner's discretization argument, we have to take the limit superior in $n$ on the left-hand side of \eqref{eq:converse}.
To ameliorate this problem, we constructed $Q^{(n)}$ in \eqref{eq:artifical}, whose third part consists of a convex combination of exponentially-weighted output distributions indexed by elements of an $\epsilon$-net in the \emph{input} probability simplex.
This differs from Tomamichel and Tan's work \cite{tomamichel_tan_2013} in which they considered an $\epsilon$-net for the \emph{output} probability simplex.
Moreover, while Tomamichel and Tan's construction of the $\epsilon$-net can yield a uniformly bounded normalization constant $F$, our construction in the third part of \eqref{eq:artifical} cannot yield a uniformly bounded normalization constant $F$ with respect to $T > 0$, but our construction yields an $F$ that scaled as $\mathrm{O}( \sqrt{T} )$ as $T \to \infty$; see~\eqref{eq:F_bound}.
This construction, and resulting unboundedness with respect to $T$, appears to be required to handle the continuous nature of Poisson channel.
Because $F = \mathrm{O}( \sqrt{T} )$ as $T \to \infty$, our upper bound on the third-order coding rate, as shown in \eqref{eq:third-order_bounds}, is $\le \log T + \mathrm{O}( 1 )$ as $T \to \infty$, which differs slightly from the lower bound which reads $\ge (1/2) \log T + \mathrm{O}( 1 )$ as $T \to \infty$.

\subsection{Technical Contributions in Proving Achievability Part of \thref{th:2nd-order}}
\label{sect:comment_direct}

The proof of the achievability part of \thref{th:2nd-order} is inspired by Polyanskiy's technique to prove the second- and third-order asymptotics for non-singular DMCs (cf.\ \cite[Section~3.4.5]{polyanskiy_thesis}).
We note, however, that Polyanskiy's proof \cite[Section~3.4.5]{polyanskiy_thesis} holds for codes without a cost constraint.
Moreover, although the (symbol-wise) discretized channel $W_{n}$ is non-singular, it depends on $n$ (see \eqref{def:W_Delta}--\eqref{def:Delta_n}).
Thus, Polyanskiy's technique cannot be adapted in a straightforward manner to yield the achievability part of \thref{th:2nd-order}.
To adapt Polyanskiy's technique to prove the third-order term for the Poisson channel, we first slightly generalize the random coding union bound so that it is amenable to handling cost constraints;
this is stated in \lemref{lem:RCU_cost}. 
By constructing an input distribution that is slightly perturbed from the CAID in \eqref{def:delta-convention} and using the concentration bound in \lemref{lem:delta-convention}, we were able to bound the probability that the cost constraint is violated.
The choice of \eqref{def:delta-convention} and \lemref{lem:delta-convention} are inspired by the delta-convention%
\footnote{The delta-convention is a technical assumption frequently used in the method of types \cite{csiszar_korner_2011} to assert that the strongly typical set is a high-probability set.} \cite[Convention~2.11]{csiszar_korner_2011} and its consequent lemma \cite[Lemma~2.12 or Problem~3.18(b)]{csiszar_korner_2011}, respectively.
Here, it is worth pointing out that if $p_{0} < \sigma$, i.e., if the weight constraint in \eqref{eq:weight_constraint} is not tight, then it suffices to employ the CAID $P^{\ast}$ (and not a perturbed version of it in~\eqref{def:delta-convention}) to generate the random code.
Next, we defined and analyzed a high-probability event $\mathcal{E}_{T}$ of \eqref{def:event_n};
this event replaces Polyanskiy's choice in \cite[Equation~(3.318)]{polyanskiy_thesis}.
In \lemref{lem:eventEn_complement}, the probability of $\mathcal{E}_{T}^{\complement}$ was shown to be appropriately bounded by appealing to a modified logarithmic Sobolev inequality (see \remref{rem:sobolev} of \appref{app:eventEn_complement}).
We showed in \lemref{lem:tight_3rd-order_term} that conditioned on $\mathcal{E}_{T}$, each of the terms in the generalized random coding union bound can be appropriately bounded.

We note that there are other techniques to show that $+(1/2)\log n +\mathrm{O}(1)$ is third-order achievable for certain channels sources; see, for example \cite{kostina_verdu_2015, tan_tomamichel_2015, moulin_2017, kosut_sankar_2017, iri_kosut_2019}.
For example, Tan and Tomamichel \cite{tan_tomamichel_2015} showed that $+ (1/2) \log n + \mathrm{O}( 1 )$ is third-order achievable for the additive white Gaussian noise channel.
However, while the key idea in \cite{tan_tomamichel_2015} is to use Laplace's approximation to bound a certain probability, in the proof of \lemref{lem:tight_3rd-order_term}  in \appref{app:tight_3rd-order_term}, we applied the Berry--Esseen theorem judiciously to bound the analogous probability.

\appendices

\section{Proof of \propref{prop:Shannon_ordering}}
\label{app:Shannon_ordering}

Given two channels $W_{1} : \mathcal{X} \to \mathcal{Y}$ and $W_{2} : \mathcal{Y} \to \mathcal{Z}$, denote by $W_{1}W_{2} : \mathcal{X} \to \mathcal{Z}$ the concatenation%
\footnote{Clearly, this concatenation operator is associative, as in the matrix multiplication.}
of $W_{1}$ and $W_{2}$, i.e.,
\begin{align}
(W_{1}W_{2})(z \mid x)
\coloneqq
\mathbb{E}[ W_{2}(z \mid Y_{x}) ]
\end{align}
for each $(x, z) \in \mathcal{X} \times \mathcal{Z}$ with a certain r.v.\ $Y_{x}$ satisfying $\mathbb{P} \circ Y_{x}^{-1} = W_{1}(\cdot \mid x)$.
Let $W_{\lambda_{0}} : \mathcal{W}(T, A, \sigma) \to \mathcal{S}( T )$ be the Poisson channel defined in \sectref{sect:preliminaries}.
By the same argument as Wyner's \emph{ad hoc} assumption \cite[Section~II in Part~I]{wyner_1988}, the discretized channel $W_{n}^{n} : \mathcal{B}(n, \sigma) \to \{ 0, 1 \}^{n}$ defined in \sectref{sect:discretization} can be seen as a concatenation of four channels
\begin{align}
W_{n}^{n}
=
U \, W_{\lambda_{0}} \, V_{1} \, V_{2} ,
\end{align}
where
the channel $U : \mathcal{B}(n, \sigma) \to \mathcal{W}(T, A, \sigma)$ is given by
\begin{align}
U(\lambda \mid x_{1}, \dots, x_{n})
=
\prod_{i=1}^{n} \bvec{1}\Big\{ \lambda(t) = x_{i} \, A \quad \forall t \in ((i-1) \, \Delta_{n}, i \, \Delta_{n}] \Big\} ,
\end{align}
the channel $V_{1} : \mathcal{S}(T) \to (\mathbb{N} \cup \{ 0 \})^{n}$ is given by
\begin{align}
V_{1}(\hat{y}_{1}, \dots, \hat{y}_{n} \mid \nu_{0}^{T})
=
\prod_{i=1}^{n} \bvec{1} \Big\{ \hat{y}_{i} = \nu( i \, \Delta_{n} ) - \nu( (i-1) \, \Delta_{n} ) \Big\} ,
\end{align}
and the channel $V_{2} : (\mathbb{N} \cup \{ 0 \})^{n} \to \{ 0, 1 \}^{n}$ is given by
\begin{align}
V_{2}(y_{1}, \dots, y_{n} \mid \hat{y}_{1}, \dots, \hat{y}_{n})
=
\prod_{i=1}^{n} \Big( \bvec{1}\big\{ y_{i} = \hat{y}_{i} \big\} + \bvec{1}\big\{ y_{i} = 0 \ \mathrm{and} \ \hat{y}_{i} \ge 2 \big\} \Big) .
\end{align}
Therefore, we can obtain \propref{prop:Shannon_ordering} by considering appropriate stochastic encoder and decoder induced by $U$ and $V_{1}V_{2}$, respectively (cf.\ \cite[Problem~6.17(b)]{csiszar_korner_2011}, \cite{shannon_1958}, \cite{nasser_2018}).
This completes the proof of \propref{prop:Shannon_ordering}.
\hfill\IEEEQEDhere

\section{Proof of \eqref{eq:asympt_Dn_gamma}}
\label{app:eq:asympt_Dn_gamma}

Denote by
\begin{align}
r_{n}^{\prime}
\coloneqq
P_{[\kappa]}^{\ast}W_{n}( 1 )
=
(1-p^{\ast}-\kappa) \, a_{n} + (p^{\ast}+\kappa) \, b_{n}
\label{def:rn_prime}
\end{align}
the parameter of the Bernoulli distribution $P_{[\kappa]}^{\ast}W_{n}$, where $a_{n}$ and $b_{n}$ are defined in \eqref{def:an} and \eqref{def:bn}, respectively.
Note that the following asymptotic equivalences hold:
\begin{align}
a_{n}
& \sim
\frac{ s \, A \, T }{ n } ,
\label{eq:asympt_an_Dn} \\
b_{n}
& \sim
\frac{ (1+s) \, A \, T }{ n } ,
\label{eq:asympt_bn_Dn} \\
r_{n}
& \sim
\frac{ (p_{n}+s) \, A \, T }{ n } ,
\label{eq:asympt_rn_Dn} \\
r_{n}^{\prime}
& \sim
\frac{ (p^{\ast}+\kappa+s) \, A \, T }{ n }
\label{eq:asympt_rn_prime}
\end{align}
as $n \to \infty$, where $r_{n}$ is defined in \eqref{def:rn_converse}.
Then, we observe that
\begin{align}
D_{n}
& \overset{\mathclap{\text{(a)}}}{=}
(1-r_{n}) \log \frac{ 1 }{ 1-r_{n}^{\ast} } + r_{n} \log \frac{ 1 }{ r_{n}^{\prime} } - (1-p_{n}) \, h( a_{n} ) - p_{n} \, h( b_{n} )
\notag \\
& \overset{\mathclap{\text{(b)}}}{=}
r_{n}^{\prime} + r_{n} \log \frac{ 1 }{ r_{n}^{\prime} } - (1-p_{n}) \, h( a_{n} ) - p_{n} \, h( b_{n} ) + \mathrm{o}( n^{-1} )
\notag \\
& \overset{\mathclap{\text{(c)}}}{=}
r_{n}^{\prime} + r_{n} \log \frac{ 1 }{ r_{n}^{\prime} } - (1-p_{n}) \, a_{n} \log \frac{ 1 }{ a_{n} } - (1-p_{n}) \, a_{n} - p_{n} \, b_{n} \log \frac{ 1 }{ b_{n} } - p_{n} \, b_{n} + \mathrm{o}( n^{-1} )
\notag \\
& \overset{\mathclap{\text{(d)}}}{=}
(r_{n}^{\prime} - r_{n}) + (1-p_{n}) \, a_{n} \log \frac{ a_{n} }{ r_{n}^{\prime} } - p_{n} \, b_{n} \log \frac{ b_{n} }{ r_{n}^{\prime} } + \mathrm{o}( n^{-1} )
\notag \\
& \overset{\mathclap{\text{(e)}}}{=}
\frac{ A \, T }{ n } \, \bigg( (p^{\ast}+\kappa-p_{n}) + (1-p_{n}) \, s \log \frac{ s }{ p^{\ast}+\kappa+s } + p_{n} \, (1+s) \log \frac{ 1+s }{ p^{\ast}+\kappa+s } \bigg) + \mathrm{o}( n^{-1} )
\label{eq:asympt_Dn_single_gamma}
\end{align}
as $n \to \infty$, where
\begin{itemize}
\item
(a) follows by \eqref{eq:Dn_gamma_type} and the definition of binary entropy function $h : u \mapsto - u \log u - (1-u) \log (1-u)$,
\item
(b) follows by the asymptotic equivalences as stated in \eqref{eq:asympt_rn_Dn}--\eqref{eq:asympt_rn_prime} and $- \log (1-u) \sim u$ as $u \to 0$,
\item
(c) follows by the asymptotic equivalences as stated in \eqref{eq:asympt_an_Dn}--\eqref{eq:asympt_bn_Dn} and $- (1-u) \log (1-u) \sim u$ as $u \to 0$,
\item
(d) follows by the definition of $r_{n} \coloneqq P_{n}W_{n}( 1 ) = (1-p_{n}) \, a_{n} + p_{n} \, b_{n}$ in \eqref{def:rn_converse}, and
\item
(e) follows by the asymptotic equivalences as stated in \eqref{eq:asympt_an_Dn}--\eqref{eq:asympt_rn_prime}.
\end{itemize}
Equation~\eqref{eq:asympt_Dn_single_gamma} indeed implies \eqref{eq:asympt_Dn_gamma}, as desired.
\hfill\IEEEQEDhere

\section{Proof of \eqref{eq:asympt_Vn_gamma}}
\label{app:eq:asympt_Vn_gamma}

Define the binary relative varentropy $v(p \, \| \, q)$ by
\begin{align}
v(p \, \| \, q)
& \coloneqq
p \, \bigg( \log \frac{ p }{ q } - d(p \, \| \, q) \bigg)^{2} + (1-p) \, \bigg( \log \frac{ 1-p }{ 1-q } - d(p \, \| \, q) \bigg)^{2} ,
\end{align}
for each $0 \le p, q \le 1$, where $d(p \, \| \, q)$ stands for the binary relative entropy defined by
\begin{align}
d(p \, \| \, q)
\coloneqq
p \log \frac{ p }{ q } + (1-p) \log \frac{ 1-p }{ 1-q } .
\label{def:binary_relative_entropy}
\end{align}
After some algebra, we obtain
\begin{align}
v(p \, \| \, q)
& =
p \, (1-p) \log^{2} \bigg( \frac{ p }{ 1-p } \frac{ 1-q }{ q } \bigg) .
\label{eq:relative_varentropy}
\end{align}
Now, a direct calculation yields
\begin{align}
V_{n}
& \overset{\mathclap{\text{(a)}}}{=}
(1-p_{n}) \, v(a_{n} \, \| \, r_{n}^{\prime}) + p_{n} \, v(b_{n} \, \| \, r_{n}^{\prime})
\notag \\
& \overset{\mathclap{\text{(b)}}}{=}
(1-p_{n}) \, a_{n} \, \log^{2} \bigg( \frac{ a_{n} }{ 1-a_{n} } \frac{ 1-r_{n}^{\prime} }{ r_{n}^{\prime} } \bigg) + p_{n} \, b_{n} \, (1-b_{n}) \log^{2} \bigg( \frac{ b_{n} }{ 1-b_{n} } \frac{ 1-r_{n}^{\prime} }{ r_{n}^{\prime} } \bigg)
\notag \\
& \overset{\mathclap{\text{(c)}}}{=}
\frac{ A \, T }{ n } \, \bigg( (1-p_{n}) \, s \log^{2} \frac{ s }{ p^{\ast}+\kappa+s } + p_{n} \, (1+s) \log^{2} \frac{ 1+s }{ p^{\ast}+\kappa+s } \bigg) + \mathrm{o}( n^{-1} )
\label{eq:asympt_Vn_app}
\end{align}
as $n \to \infty$, where
\begin{itemize}
\item
(a) follows from \eqref{eq:Vn_gamma_type} and the definition of $r_{n}^{\prime}$ in \eqref{def:rn_prime},
\item
(b) follows from \eqref{eq:relative_varentropy}, and
\item
(c) follows from the asymptotic equivalences as stated in \eqref{eq:asympt_an_Dn}--\eqref{eq:asympt_rn_prime}.
\end{itemize}

We now employ the following inequality%
\footnote{\lemref{lem:q-log} is a generalization of the well-known information theoretic inequalities: $1 - x^{-1} \le \log x \le x - 1$.}.

\begin{lemma}[{\cite[Lemma~1]{sakai_iwata_isit2016}}]
\label{lem:q-log}
For every $u > 0$ and $q > r$, it holds that $\ln_{q} u \le \ln_{r} u$ with equality if and only if $u = 1$, where $\ln_{q} : (0, \infty) \to \mathbb{R}$ stands for the $q$-logarithm function \cite{tsallis_1994} defined by
\begin{align}
\ln_{q} u
\coloneqq
\begin{dcases}
\frac{ u^{1-q} - 1 }{ 1-q }
& \mathrm{if} \ q \neq 1 ,
\\
\log u
& \mathrm{if} \ q = 1 .
\end{dcases}
\end{align}
\end{lemma}

It follows from \lemref{lem:q-log} with $q = 1$ and $r = 1/2$ that
\begin{align}
\log u
\le
2 \, (\sqrt{ u } - 1)
<
2 \, \sqrt{ u }
\label{eq:q-log_half}
\end{align}
for every $u > 0$.
Now, we have
\begin{align}
\limsup_{n \to \infty} n \, V_{n}
& \overset{\mathclap{\text{(a)}}}{=}
A \, T \limsup_{n \to \infty} \bigg( (1-p_{n}) \, s \log^{2} \frac{ s }{ p^{\ast}+\kappa+s } + p_{n} \, (1+s) \log^{2} \frac{ 1+s }{ p^{\ast}+\kappa+s } \bigg)
\notag \\
& <
A \, T \, \bigg( s \log^{2} \frac{ s }{ p^{\ast}+\kappa+s } + (1+s) \log^{2} \frac{ 1+s }{ p^{\ast}+\kappa+s } \bigg)
\notag \\
& \overset{\mathclap{\text{(b)}}}{<}
4 \, A \, T \, \bigg( s \, \bigg( \frac{ p^{\ast}+\kappa+s }{ s } \bigg) + (1+s) \, \bigg( \frac{ p^{\ast}+\kappa+s }{ 1+s } \bigg) \bigg)
\notag \\
& =
8 \, A \, T \, (p^{\ast} + \kappa + s) ,
\end{align}
where
\begin{itemize}
\item
(a) follows from \eqref{eq:asympt_Vn_app}, and
\item
(b) follows from \eqref{eq:q-log_half}.
\end{itemize}
This completes the proof of \eqref{eq:asympt_Vn_gamma}.
\hfill\IEEEQEDhere

\section{Proof of \eqref{eq:Dk_quadratic}}
\label{app:Dk_quadratic}

After some algebra, we observe that
\begin{align}
D(W_{n_{k}} \, \| \, \tilde{P}_{k}W_{n_{k}} \mid P_{n_{k}})
& =
I(P_{n_{k}}, W_{n_{k}}) + D(P_{n_{k}}W_{n_{k}} \, \| \, \tilde{P}_{k}W_{n_{k}})
\label{eq:algebra_log}
\end{align}
for each $k \ge 1$, and one has%
\footnote{While Tomamichel and Tan \cite[Property~4 of Lemma~7]{tomamichel_tan_2013} gave an upper bound on the relative entropy $D(P \, \| \, Q)$ by the reverse Pinsker inequality $\le |P - Q|^{2}/Q_{\min}$ to prove \cite[Proposition~8]{tomamichel_tan_2013}, we cannot get a useful upper bound by the reverse Pinsker inequality to prove the second-order converse of the Poisson channel, because for the Poisson channel, it holds that $Q_{\min} = \Theta( n_{k}^{-1} )$ as $k \to \infty$.}
\begin{align}
D(P_{n_{k}}W_{n_{k}} \, \| \, \tilde{P}_{k}W_{n_{k}})
& \overset{\mathclap{\text{(a)}}}{=}
r_{n_{k}} \log \frac{ r_{n_{k}} }{ \tilde{r}_{k} } + (1 - r_{n_{k}}) \log \frac{ 1 - r_{n_{k}} }{ 1 - \tilde{r}_{k} }
\notag \\
& \overset{\mathclap{\text{(b)}}}{=}
\frac{ (p_{n_{k}}+s) \, A \, T }{ n_{k} } \log \frac{ p_{n_{k}}+s }{ \tilde{p}_{k}+s } + (1 - r_{n_{k}}) \log \frac{ 1 - r_{n_{k}} }{ 1 - \tilde{r}_{k} } + \mathrm{o}( n_{k}^{-1} )
\notag \\
& \overset{\mathclap{\text{(c)}}}{\le}
(1 - r_{n_{k}}) \log \frac{ 1 - r_{n_{k}} }{ 1 - \tilde{r}_{k} } + \mathrm{o}( n_{k}^{-1} )
\notag \\
& \le
\log \frac{ 1 - r_{n_{k}} }{ 1 - \tilde{r}_{k} } + \mathrm{o}( n_{k}^{-1} )
\notag \\
& \overset{\mathclap{\text{(d)}}}{=}
\frac{ A \, T }{ n_{k} } \Big( \tilde{p}_{k} - p_{n_{k}} \Big) + \mathrm{o}( n_{k}^{-1} )
\notag \\
& \overset{\mathclap{\text{(e)}}}{=}
\frac{ A \, T }{ n_{k} } \bigg( p^{\ast} + \frac{ m_{k} }{ T } - p_{n_{k}} \bigg) + \mathrm{o}( n_{k}^{-1} )
\notag \\
& \overset{\mathclap{\text{(f)}}}{\le}
\frac{ A \, T }{ n_{k} } \bigg( p^{\ast} + \frac{ m_{k} }{ T } - p^{\ast} - \frac{ m_{k}-1 }{ T } \bigg) + \mathrm{o}( n_{k}^{-1} )
\notag \\
& =
\frac{ A }{ n_{k} } + \mathrm{o}( n_{k}^{-1} )
\label{eq:instead_reversePinsker}
\end{align}
as $k \to \infty$, where
\begin{itemize}
\item
(a) follows from the definitions of $r_{n_{k}}$ and $\tilde{r}_{k}$ in \eqref{def:rn_converse} and \eqref{def:rk_tilde}, respectively,
\item
(b) follows from the facts that
\begin{align}
r_{n}
& \sim
\frac{ (p_{n}+s) \, A \, T }{ n }
\qquad (\mathrm{as} \ n \to \infty) ,
\label{eq:asympt_rn} \\
\tilde{r}_{k}
& \sim
\frac{ (\tilde{p}_{k}+s) \, A \, T }{ n_{k} }
\qquad (\mathrm{as} \ k \to \infty) ,
\label{eq:asympt_rk}
\end{align}
\item
(c) follows from the definitions of $\tilde{m}_{T}$ and $\tilde{p}_{k}$ in \eqref{def:mk} and \eqref{def:p_tilde}, respectively, implying that $p_{n_{k}} \le \tilde{p}_{k}$ or
\begin{align}
\log \frac{ p_{n_{k}}+s }{ \tilde{p}_{k}+s }
\le
0 ,
\end{align}
\item
(d) follows from \eqref{eq:asympt_rn}--\eqref{eq:asympt_rk} and the fact that
\begin{align}
\log(1 - u)
\sim
- u
\qquad (\mathrm{as} \ u \to 0) ,
\end{align}
\item
(e) follows by the definition of $\tilde{p}_{k}$ in \eqref{def:p_tilde}, and
\item
(f) follows by the definition of $m_{k}$ in \eqref{def:mk}.
\end{itemize}
On the other hand, one can see in the same way as \appref{app:eq:asympt_Dn_gamma} that
\begin{align}
I(P_{n_{k}}, W_{n_{k}})
=
\frac{ T }{ n_{k} } \, g( p_{n_{k}} ) + \mathrm{o}( n_{k}^{-1} )
\label{eq:mutual}
\end{align}
as $k \to \infty$, where the function $g : [0, 1] \to \mathbb{R}$ is defined by
\begin{align}
g( u )
\coloneqq
A \, \bigg( (1-u) \, s \log \frac{ s }{ u+s } + u \, (1+s) \log \frac{ 1+s }{ u+s } \bigg) .
\label{def:g}
\end{align}
Direct calculations show
\begin{align}
\frac{ \mathrm{d} }{ \mathrm{d} u } g( u )
& =
A \, \Bigg( s \, \bigg( - \log s + s \log (u+s) - \frac{ 1-u }{ u+s } \bigg) + (1+s) \, \bigg( \log (1+s) - \log (u+s) - \frac{ u }{ u+s } \bigg) \Bigg)
\notag \\
& =
A \log \bigg( \frac{ (1+s)^{1+s} }{ s^{s} \, \mathrm{e} } \frac{ 1 }{ u+s } \bigg)
\notag \\
& =
A \log \frac{ p_{0}+s }{ u+s } ,
\\
\frac{ \mathrm{d}^{2} }{ \mathrm{d} u^{2} } g( u )
& =
- \frac{ A }{ u+s } ,
\\
\frac{ \mathrm{d}^{3} }{ \mathrm{d} u^{3} } g( u )
& =
\frac{ A }{ (u+s)^{2} } ,
\\
\frac{ \mathrm{d}^{4} }{ \mathrm{d} u^{4} } g( u )
& =
- \frac{ 2 \, A }{ (u+s)^{3} } ,
\end{align}
where $p_{0}$ is defined in \eqref{def:p0}.
Now, it follows from \eqref{def:p_ast} that%
\footnote{The first-order derivative in \eqref{eq:first-order-derivative_C} is zero for the usual discrete memoryless channels (DMCs) without cost-constraint. However, the first-order derivative in \eqref{eq:first-order-derivative_C} can be positive when there is a cost constraint on the codewords.}
\begin{align}
\frac{ \mathrm{d} }{ \mathrm{d} u } g( u ) \bigg|_{u = p^{\ast}}
=
A \log \frac{ p_{0}+s }{ p^{\ast}+s }
& =
\begin{dcases}
= 0
& \mathrm{if} \ p_{0} \le \sigma ,
\\
> 0
& \mathrm{if} \ p_{0} > \sigma ,
\end{dcases}
\label{eq:first-order-derivative_C}
\end{align}
and it follows from \eqref{def:p0}, \eqref{def:xn}, and \eqref{def:pn} that $p_{n_{k}} < p^{\ast}$ if $p_{0} > \sigma$.
Therefore, it follows by Taylor's theorem applied to $g( \cdot )$ around $p^{\ast}$ that there exists a real number $\hat{p}$ between $p^{\ast}$ and $p_{n_{k}}$ such that%
\footnote{The upper bound \eqref{eq:g_Taylor} is a counterpart of \cite[Property~3 in Lemma~7]{tomamichel_tan_2013}.}
\begin{align}
g( p_{n_{k}} )
& =
g( p^{\ast} ) + \bigg( \frac{ \mathrm{d} }{ \mathrm{d} u } g( u ) \bigg|_{u = p^{\ast}} \bigg) \, (p_{n_{k}} - p^{\ast}) + \bigg( \frac{ \mathrm{d}^{2} }{ \mathrm{d} u^{2} } g( u ) \bigg|_{u = p^{\ast}} \bigg) \, \frac{ (p_{n_{k}} - p^{\ast})^{2} }{ 2 }
\notag \\
& \qquad \qquad \qquad {}
+ \bigg( \frac{ \mathrm{d}^{3} }{ \mathrm{d} u^{3} } g( u ) \bigg|_{u = p^{\ast}} \bigg) \, \frac{ (p_{n_{k}} - p^{\ast})^{3} }{ 6 } + \bigg( \frac{ \mathrm{d}^{4} }{ \mathrm{d} u^{4} } g( u ) \bigg|_{u = \hat{p}} \bigg) \, \frac{ (p_{n_{k}} - p^{\ast})^{4} }{ 24 }
\notag \\
& =
g( p^{\ast} ) - |p_{n_{k}} - p^{\ast}| \, A \log \frac{ p_{0}+s }{ p^{\ast}+s } - \frac{ A \, (p_{n_{k}} - p^{\ast})^{2} }{ 2 \, (p^{\ast}+s) } + \frac{ A \, (p_{n_{k}} - p^{\ast})^{3} }{ 6 \, (p^{\ast}+s)^{2} } - \frac{ A \, (p_{n_{k}} - p^{\ast})^{4} }{ 12 \, (\hat{p}+s)^{3} }
\notag \\
& \le
g( p^{\ast} ) - |p_{n_{k}} - p^{\ast}| \, A \log \frac{ p_{0}+s }{ p^{\ast}+s } - \frac{ A \, (p_{n_{k}} - p^{\ast})^{2} }{ 2 \, (p^{\ast}+s) } + \frac{ A \, (p_{n_{k}} - p^{\ast})^{3} }{ 6 \, (p^{\ast}+s)^{2} }
\notag \\
& =
g( p^{\ast} ) - |p_{n_{k}} - p^{\ast}| \, A \log \frac{ p_{0}+s }{ p^{\ast}+s } - \frac{ A \, (p_{n_{k}} - p^{\ast})^{2} }{ 2 \, (p^{\ast}+s) } \bigg( 1 - \frac{ p_{n_{k}}-p^{\ast} }{ 3 \, (p^{\ast}+s) } \bigg)
\notag \\
& \overset{\mathclap{\text{(a)}}}{\le}
g( p^{\ast} ) - (p_{n_{k}} - p^{\ast})^{2} \, A \log \frac{ p_{0}+s }{ p^{\ast}+s } - \frac{ A \, (p_{n_{k}} - p^{\ast})^{2} }{ 2 \, (p^{\ast}+s) } \bigg( 1 - \frac{ p_{n_{k}}-p^{\ast} }{ 3 \, (p^{\ast}+s) } \bigg)
\notag \\
& \overset{\mathclap{\text{(b)}}}{\le}
g( p^{\ast} ) - (p_{n_{k}} - p^{\ast})^{2} \, A \log \frac{ p_{0}+s }{ p^{\ast}+s } - \frac{ A \, (p_{n_{k}} - p^{\ast})^{2} }{ 2 \, (p^{\ast}+s) } \bigg( 1 - \frac{ \kappa }{ 3 \, (p^{\ast}+s) } \bigg)
\label{eq:g_Taylor}
\end{align}
for each $k \ge 1$, where
\begin{itemize}
\item
(a) follows from \eqref{eq:first-order-derivative_C} and the fact that
\begin{align}
(p_{n_{k}} - p^{\ast})^{2}
<
|p_{n_{k}} - p^{\ast}|
<
1 ,
\end{align}
and
\item
(b) follows by the definition of $\mathcal{I}_{3} = \{ n_{k} \}_{k = 1}^{\infty}$ in \eqref{eq:I3}.
\end{itemize}
Note from \eqref{eq:first-order-derivative_C} that the second term
\begin{align}
- (p_{n_{k}} - p^{\ast})^{2} \, A \log \frac{ p_{0}+s }{ p^{\ast}+s }
\end{align}
in \eqref{eq:g_Taylor} is \emph{zero} if $p_{0} \le \sigma$, and is \emph{negative} if $p_{0} > \sigma$.
In any case, we have from \eqref{eq:g_Taylor} that $g( p_{n_{k}} ) \le g( p^{\ast} ) - (\text{positive const.}) \times (p_{n_{k}} - p^{\ast})^{2}$.
By the definitions of $C^{\ast}$ and $g( \cdot )$ in \eqref{def:capacity_Poisson} and \eqref{def:g}, respectively, it is clear that
\begin{align}
g( p^{\ast} )
=
C^{\ast} .
\label{eq:g_Cast}
\end{align}
Hence, we obtain
\begin{align}
&
n_{k} \, D(W_{n_{k}} \, \| \, \tilde{P}_{k}W_{n_{k}} \mid P_{n_{k}})
\notag \\
& \quad \overset{\mathclap{\text{(a)}}}{=}
n_{k} \, I(P_{n_{k}}, W_{n_{k}}) + n_{k} \, D(P_{n_{k}}W_{n_{k}} \, \| \, \tilde{P}_{k}W_{n_{k}})
\notag \\
& \quad \overset{\mathclap{\text{(b)}}}{\le}
n_{k} \, I(P_{n_{k}}, W_{n_{k}}) + A + \mathrm{o}( 1 )
\notag \\
& \quad \overset{\mathclap{\text{(c)}}}{\le}
T \, C^{\ast} - T \, (p_{n_{k}} - p^{\ast})^{2} \, A \, \Bigg( \log \frac{ p_{0}+s }{ p^{\ast}+s } + \frac{ 1 }{ 2 \, (p^{\ast}+s) } \bigg( 1 - \frac{ \kappa }{ 3 \, (p^{\ast}+s) } \bigg) \Bigg) + A + \mathrm{o}( 1 )
\label{eq:bound_D}
\end{align}
as $k \to \infty$, where
\begin{itemize}
\item
(a) follows from \eqref{eq:algebra_log},
\item
(b) follows from \eqref{eq:instead_reversePinsker}, and
\item
(c) follows from \eqref{eq:mutual}, \eqref{eq:g_Taylor}, and \eqref{eq:g_Cast}.
\end{itemize}
Here, note that the terms in parentheses in the second term in \eqref{eq:bound_D} is strictly positive, i.e.,
\begin{align}
\log \frac{ p_{0}+s }{ p^{\ast}+s } + \frac{ 1 }{ 2 \, (p^{\ast}+s) } \bigg( 1 - \frac{ \kappa }{ 3 \, (p^{\ast}+s) } \bigg)
>
0 ,
\label{eq:positivity}
\end{align}
because of \eqref{eq:bounds_p_ast}, \eqref{def:kappa}, and \eqref{eq:first-order-derivative_C}.
This completes the proof of \eqref{eq:Dk_quadratic} with the identification of $G_{1}$ as the constant in \eqref{eq:positivity}.
\hfill\IEEEQEDhere

\section{Proof of \eqref{eq:Vk_quadratic}}
\label{app:Vk_quadratic}

Similar to \eqref{eq:asympt_Vn_app}, we may observe that
\begin{align}
n_{k} \, V(W_{n_{k}} \, \| \, \tilde{P}_{k}W_{n_{k}} \mid P_{n_{k}})
=
T \, \tilde{v}( p_{n_{k}}, \tilde{p}_{k} ) + \mathrm{o}( 1 )
\label{eq:Vk_asympt}
\end{align}
as $k \to \infty$, where the mapping $\tilde{v} : [0, 1] \times [0, 1] \to \mathbb{R}$ is defined by
\begin{align}
\tilde{v}(t, u)
\coloneqq
A \, \bigg( (1-t) \, s \log^{2} \frac{ s }{ u+s } + t \, (1+s) \log^{2} \frac{ 1+s }{ u+s } \bigg) .
\label{def:v_tilde}
\end{align}
Since $\sqrt{ \tilde{v}(t, u) }$ is a continuously differentiable function of $u \in (0, 1]$ for each fixed $t \in [0, 1]$, and since $\tilde{p}_{k}$ and $p_{n_{k}}$ are bounded away from zero for all $k \ge 1$, it follows by the Lipschitz continuity%
\footnote{Note that every continuously differentiable function is Lipschitz continuous and the Lipschitz constant can be taken to be the supremum of the absolute value of the first derivative (over its domain).}
of $u \mapsto \sqrt{ \tilde{v}(t, u) }$ that there exists a Lipschitz constant $\beta_{1}( t ) > 0$ satisfying
\begin{align}
\Big| \sqrt{ \tilde{v}( t, \tilde{p}_{k} ) } - \sqrt{ \tilde{v}( t, p_{n_{k}} ) } \Big|
& \le
\beta_{1}( t ) \, |\tilde{p}_{k} - p_{n_{k}}|
\notag \\
& \overset{\mathclap{\text{(b)}}}{=}
\beta_{1}( t ) \, \bigg( p^{\ast} + \frac{ m_{k} }{ T } - p_{n_{k}} \bigg)
\notag \\
& \overset{\mathclap{\text{(c)}}}{\le}
\beta_{1}( t ) \, \bigg( p^{\ast} + \frac{ m_{k} }{ T } - p^{\ast} - \frac{ m_{k} - 1 }{ T } \bigg)
\notag \\
& =
\frac{ \beta_{1}( t ) }{ T }
\label{eq:Lipschitz_1a}
\end{align}
for every $k \ge 1$ and $0 \le t \le 1$, where
\begin{itemize}
\item
(a) follows by the definition of $\tilde{p}_{k}$ in \eqref{def:p_tilde}, and
\item
(b) follows by the definition of $m_{k}$ in \eqref{def:mk}.
\end{itemize}
Here, it can be verified by a direct calculation of the partial derivative of $u \mapsto \sqrt{ \tilde{v}(t, u) }$ with fixed $t$ that the Lipschitz constant $\beta_{1}( t )$ is continuous in $t \in [0, 1]$.
Moreover, as $p_{n_{k}}$ belongs to the closed interval $[p^{\ast} - \kappa, p^{\ast} + \kappa]$ (see \eqref{eq:I3}), inequality~\eqref{eq:Lipschitz_1a} can be uniformly relaxed as
\begin{align}
\Big| \sqrt{ \tilde{v}(p_{n_{k}}, \tilde{p}_{k}) } - \sqrt{ \tilde{v}(p_{n_{k}}, p_{n_{k}}) } \Big|
\le
\frac{ \beta_{1} }{ T }
\label{eq:Lipschitz_1b}
\end{align}
for every $k \ge 1$, where the absolute constant $\beta_{1} > 0$ is given by
\begin{align}
\beta_{1}
\coloneqq
\max_{t \in [p^{\ast}-\kappa, p^{\ast}+\kappa]} \beta_{1}( t ) .
\end{align}
Analogously, since $\sqrt{ \tilde{v}(t, t) }$ is a continuously differentiable function of $t \in (0, 1]$, it follows by the Lipschitz continuity of $t \mapsto \sqrt{ \tilde{v}(t, t) }$ that there exists an absolute constant $\beta_{2} > 0$ satisfying
\begin{align}
\Big| \sqrt{ \tilde{v}( p_{n_{k}}, p_{n_{k}} ) } - \sqrt{ \tilde{v}( p^{\ast}, p^{\ast} ) } \Big|
\le
\beta_{2} \, |p_{n_{k}} - p^{\ast}|
\label{eq:Lipschitz_2}
\end{align}
for every $k \ge 1$.
By the definitions of $V^{\ast}$ and $\tilde{v}(\cdot, \cdot)$ in \eqref{def:dispersion_Poisson} and \eqref{def:v_tilde}, respectively, it is clear that
\begin{align}
\tilde{v}( p^{\ast}, p^{\ast} )
=
V^{\ast}
\label{eq:v_tilde_V_ast}
\end{align}
Combining \eqref{eq:Vk_asympt}, \eqref{eq:Lipschitz_1a}, \eqref{eq:Lipschitz_2}, and \eqref{eq:v_tilde_V_ast}, it follows by the triangle inequality that
\begin{align}
\Big| \sqrt{ n_{k} \, V(W_{n_{k}} \, \| \, \tilde{P}_{k}W_{n_{k}} \mid P_{n_{k}}) } - \sqrt{ T \, V^{\ast} } \Big|
\le
\frac{ \beta_{1} }{ \sqrt{T} } + \sqrt{T} \, |p_{n_{k}} - p^{\ast}| \, \beta_{2} + \mathrm{o}( 1 )
\end{align}
as $k \to \infty$.
This completes the proof of \eqref{eq:Vk_quadratic}.
\hfill\IEEEQEDhere

\section{Proof of \eqref{eq:Taylor_Phi_inverse}}
\label{app:Taylor_Phi_inverse}

Since $(t, u) \mapsto \sqrt{ \tilde{v}(t, u) }$ is continuous and positive on $(0, 1) \times (0, 1)$, and since $p_{n_{k}}$ and $\tilde{p}_{k}$ are in the closed intervals $[p^{\ast} - \kappa, p^{\ast} + \kappa]$ and $[p^{\ast}-\kappa, p^{\ast}+2 \kappa]$, respectively (see \eqref{eq:I3}, \eqref{def:mk}, \eqref{eq:bound_mk}, and \eqref{def:p_tilde}), it follows by the extreme value theorem that there exists a constant $v_{\min} > 0$ satisfying
\begin{align}
\sqrt{ v(p_{n_{k}}, \tilde{p}_{k}) }
\ge
v_{\min}
\end{align}
for every $k \ge 1$ yielding together with \eqref{eq:Vk_asympt} that
\begin{align}
n_{k} \, V(W_{n_{k}} \, \| \, \tilde{P}_{k}W_{n_{k}} \mid P_{n_{k}})
\ge
T \, v_{\min} + \zeta_{k}
\label{eq:v_min}
\end{align}
for some $\zeta = \mathrm{o}( 1 )$ as $k \to \infty$.

Now, define the binary absolute and central third-moment divergence by
\begin{align}
\xi(p \, \| \, q)
& \coloneqq
p \, \bigg| \log \frac{ p }{ q } - d(p \, \| \, q) \bigg|^{3} + (1-p) \, \bigg| \log \frac{ 1-p }{ 1-q } - d(p \, \| \, q) \bigg|^{3}
\end{align}
for each $0 \le p, q \le 1$ and where $d(p \| \, q)$ is defined in \eqref{def:binary_relative_entropy}.
Similar to \eqref{eq:relative_varentropy}, we see that
\begin{align}
\xi(p \, \| \, q)
=
p \, (1-p) \, (p^{2} + (1-p)^{2}) \, \bigg| \log^{3} \bigg( \frac{ p }{ 1-p } \frac{ 1-q }{ q } \bigg) \bigg| .
\label{eq:relative_third-moment}
\end{align}
Then, we obtain
\begin{align}
\Xi_{k}
& \overset{\mathclap{\text{(a)}}}{=}
(1-p_{n_{k}}) \, \xi(a_{n_{k}} \, \| \, \tilde{r}_{k}) + p_{n_{k}} \, \xi(b_{n_{k}} \, \| \, \tilde{r}_{k})
\notag \\
& \overset{\mathclap{\text{(b)}}}{=}
(1-p_{n_{k}}) \, a_{n_{k}} \, (1-a_{n_{k}}) \, (a_{n_{k}}^{2} + (1-a_{n_{k}})^{2}) \, \bigg| \log^{3} \bigg( \frac{ a_{n_{k}} }{ 1-a_{n_{k}} } \frac{ 1-\tilde{r}_{k} }{ \tilde{r}_{k} } \bigg) \bigg|
\notag \\
& \qquad \qquad {}
+ p_{n_{k}} \, b_{n_{k}} \, (1-b_{n_{k}}) \, (b_{n_{k}}^{2} + (1-b_{n_{k}})^{2}) \, \bigg| \log^{3} \bigg( \frac{ b_{n_{k}} }{ 1-b_{n_{k}} } \frac{ 1-\tilde{r}_{k} }{ \tilde{r}_{k} } \bigg) \bigg|
\notag \\
& \overset{\mathclap{\text{(c)}}}{=}
\frac{ A \, T }{ n_{k} } \, \bigg( - (1-p_{n_{k}}) \, s \log^{3} \frac{ s }{ \tilde{p}_{k}+s } + p_{n_{k}} \, (1+s) \log^{3} \frac{ 1+s }{ \tilde{p}_{k}+s } \bigg) + \mathrm{o}( n_{k}^{-1} )
\end{align}
as $k \to \infty$, where
\begin{itemize}
\item
(a) follows from \eqref{eq:Tn_type_ast},
\item
(b) follows from \eqref{eq:relative_third-moment}, and
\item
(c) follows from the asymptotic equivalences as stated in \eqref{eq:asympt_an_Dn}--\eqref{eq:asympt_rn_Dn} and \eqref{eq:asympt_rk}.
\end{itemize}
Similar to \eqref{eq:v_min}, it follows by the extreme value theorem that there exists a constant $\xi_{\max} > 0$ satisfying
\begin{align}
n_{k} \, \Xi(W_{n_{k}} \, \| \, \tilde{P}_{k}W_{n_{k}} \mid P_{n_{k}})
\le
T \, \xi_{\max} + \mathrm{o}( 1 )
\label{eq:xi_max}
\end{align}
as $k \to \infty$.
Combining \eqref{eq:v_min} and \eqref{eq:xi_max}, there exists a positive sequence $\delta_{k} = \mathrm{o}( 1 )$ (as $k \to \infty$) satisfying
\begin{align}
\frac{ 6 \, \Xi(W_{n_{k}} \, \| \, \tilde{P}_{k}W_{n_{k}} \mid P_{n_{k}}) }{ \sqrt{ n_{k} \, V(W_{n_{k}} \, \| \, \tilde{P}_{k}W_{n_{k}} \mid P_{n_{k}}) } }
\le
\frac{ 6 \, \xi_{\max} }{ \sqrt{ T \, v_{\min}^{3} } } + \delta_{k} .
\end{align}
Supposing that $\eta = 1/\sqrt{T}$, since
\begin{align}
\epsilon_{n_{k}} + \delta_{k} + \frac{ 1 }{ \sqrt{T} } + \frac{ 6 \, \xi_{\max} }{ \sqrt{ T \, v_{\min}^{3} } }
<
\frac{ 1 - \varepsilon  }{ 2 }
\end{align}
for sufficiently large $k$ and $T$, it follows from a Taylor series expansion of $\Phi^{-1}( \cdot )$ around $\varepsilon + \epsilon_{n_{k}} + \delta_{k}$ that
\begin{align}
\Phi^{-1}\left( \varepsilon + \epsilon_{n_{k}} + \delta_{k} + \frac{ 1 }{ \sqrt{T} } + \frac{ 6 \, \xi_{\max} }{ \sqrt{ T \, v_{\min}^{3} } } \right)
\le
\Phi^{-1}(\varepsilon + \epsilon_{n_{k}} + \delta_{k}) + \frac{ v_{\min}^{3/2} + 6 \, \xi_{\max} }{ v_{\min}^{3/2} \sqrt{ T }} \, \tilde{G}_{2}
\end{align}
for sufficiently large $k$ and $T$, where the positive constant $\tilde{G}_{2}$ is given as
\begin{align}
\tilde{G}_{2}
=
\sqrt{ 2 \pi } \max\left\{ \exp\left( \frac{ 1 }{ 2 } \, \Phi^{-1}( \varepsilon )^{2} \right), \exp\left( \frac{ 1 }{ 2 } \, \Phi^{-1}\left( \frac{ 1 + \varepsilon }{ 2 } \right)^{2}\right) \right\}
\end{align}
depending only on the tolerated probability of error $0 < \varepsilon < 1$.
This completes the proof of \eqref{eq:Taylor_Phi_inverse}.
\hfill\IEEEQEDhere

\section{Proof of \lemref{lem:RCU_cost}}
\label{app:RCU_cost}

Let $X_{1}, X_{2}, \dots, X_{M}$ be i.i.d.\ r.v.'s with generic distribution $P$.
Consider a maximum-likelihood decoder $g_{\mathrm{ML}} : \mathcal{Y} \to \mathcal{X}$ satisfying
\begin{align}
g_{\mathrm{ML}}( Y )
\in
\argmax\limits_{m \in \{ 1, \dots, M \}} W(Y \mid X_{m})
\end{align}
with probability $1$.
Then, the error probability averaged over the ensemble of random codes $\{ X_{m} \}_{m=1}^{M}$ is given by
\begin{align}
\mathbb{E}\bigg[ \frac{ 1 }{ M } \sum_{m = 1}^{M} \mathbb{P}\{ \chi(X) > \beta \ \mathrm{or} \ g_{\mathrm{ML}}( Y ) \neq X \mid X = X_{m} \} \bigg]
& \le
\mathbb{E}\bigg[ \frac{ 1 }{ M } \sum_{m = 1}^{M} \mathbb{P}\{ \chi(X) > \beta \mid X = X_{m} \} + \frac{ 1 }{ M } \sum_{m=1}^{M} \mathbb{P}\{ g_{\mathrm{ML}}( Y ) \neq X \mid X = X_{m} \} \bigg]
\notag \\
& =
\mathbb{P}\{ \chi(X) > \beta \} + \mathbb{E}[ \mathbb{P}\{ g_{\mathrm{ML}}( Y ) \neq X \mid X = X_{1} \} ] .
\label{eq:union_cost}
\end{align}
By the standard argument of the random coding union bound (see the proof of \cite[Theorem~16]{polyanskiy_poor_verdu_2010}), the last term in the right-hand side on \eqref{eq:union_cost} can be bounded from above by
\begin{align}
\mathbb{E}[ \mathbb{P}\{ g_{\mathrm{ML}}( Y ) \neq X \mid X = X_{1} \} ]
& \le
\mathbb{E}\Bigg[ \min\bigg\{ 1, (M-1) \, \mathbb{P}\bigg\{ \log \frac{ W(Y \mid \bar{X}) }{ PW( Y ) } \ge \log \frac{ W(Y \mid X) }{ PW( Y ) } \ \bigg| \ X, Y \bigg\} \bigg\} \Bigg] ,
\end{align}
which asserts \lemref{lem:RCU_cost} together with \eqref{eq:union_cost}.
\hfill\IEEEQEDhere

\section{Proof of \lemref{lem:unconditional_divergences}}
\label{app:unconditional_divergences}

Since $n = \omega( T )$ as $T \to \infty$ (see \eqref{def:n_omegaT}), it follows that $a_{n} = \mathrm{o}( 1 )$ and $b_{n} = \mathrm{o}( 1 )$ as $T \to \infty$.
We readily see that
\begin{align}
I(P_{n}, W_{n})
& =
h( r_{n} ) - (1-p_{n}) \, h( a_{n} ) - p_{n} \, h( b_{n} )
\notag \\
& \overset{\mathclap{\text{(a)}}}{=}
r_{n} \log \frac{ 1 }{ r_{n} } - (1-p_{n}) \, a_{n} \log \frac{ 1 }{ a_{n} } - p_{n} \, b_{n} \log \frac{ 1 }{ b_{n} } + \mathrm{o}( n^{-1} )
\notag \\
& \overset{\mathclap{\text{(b)}}}{=}
(1-p_{n}) \, a_{n} \log \frac{ a_{n} }{ r_{n} } + p_{n} \, b_{n} \log \frac{ b_{n} }{ r_{n} } + \mathrm{o}( n^{-1} )
\notag \\
& \overset{\mathclap{\text{(c)}}}{=}
(1-p^{\ast}) \, a_{n} \log \frac{ s }{ p^{\ast}+s } + p^{\ast} \, b_{n} \log \frac{ 1+s }{ p^{\ast}+s } + \mathrm{o}( n^{-1} )
\notag \\
& \overset{\mathclap{\text{(d)}}}{=}
\frac{ A \, T }{ n } \bigg( (1-p^{\ast}) \, s \log \frac{ s }{ p^{\ast}+s } + p^{\ast} \, (1+s) \log \frac{ 1+s }{ p^{\ast}+s } \bigg) + \mathrm{o}( n^{-1} )
\notag \\
& =
\frac{ T \, C^{\ast} }{ n } + \mathrm{o}( n^{-1} )
\label{eq:asympt_I}
\end{align}
as $T \to \infty$, where
\begin{itemize}
\item
(a) follows from the fact that $- (1-u) \log (1-u) \sim u$ as $u \to 0$,
\item
(b) follows by the definition of $r_{n} = (1-p_{n}) \, a_{n} + p_{n} \, b_{n}$ in \eqref{def:output_rn_direct},
\item
(c) follows from the facts that
\begin{align}
\lim_{T \to \infty} p_{n}
& =
p^{\ast} ,
\\
\lim_{T \to \infty} \frac{ a_{n} }{ r_{n} }
& =
\frac{ s }{ p^{\ast}+s } ,
\\
\lim_{T \to \infty} \frac{ b_{n} }{ r_{n} }
& =
\frac{ 1+s }{ p^{\ast}+s } ,
\end{align}
and
\item
(d) follows from the asymptotic equivalences as stated in \eqref{eq:asympt_an_Dn}--\eqref{eq:asympt_bn_Dn}.
\end{itemize}
Equation~\eqref{eq:asympt_I} implies \eqref{eq:asympt_I_direct} of \lemref{lem:unconditional_divergences}.

We shall next verify \eqref{eq:asympt_V_direct} of \lemref{lem:unconditional_divergences}.
Noting the asymptotic equivalences stated in \eqref{eq:asympt_an_Dn}--\eqref{eq:asympt_bn_Dn} and
\begin{align}
n \, r_{n}
& \sim
(p^{\ast}+s) \, A \, T
\qquad (\mathrm{as} \ T \to \infty) ,
\\
u
& \sim
- \log (1-u)
\qquad (\mathrm{as} \ u \to 0) ,
\end{align}
it follows from \eqref{eq:asympt_I} that
\begin{align}
\bigg( \log \frac{ 1-a_{n} }{ 1-r_{n} } - I(P_{n}, W_{n}) \bigg)^{2}
& =
\mathrm{O}( n^{-2} ) ,
\\
\bigg( \log \frac{ 1-b_{n} }{ 1-b_{n} } - I(P_{n}, W_{n}) \bigg)^{2}
& =
\mathrm{O}( n^{-2} ) ,
\\
\bigg( \log \frac{ a_{n} }{ r_{n} } - I(P_{n}, W_{n}) \bigg)^{2}
& =
\log^{2} \frac{ s }{ p^{\ast}+s } + \mathrm{O}( n^{-1} ) ,
\\
\bigg( \log \frac{ b_{n} }{ r_{n} } - I(P_{n}, W_{n}) \bigg)^{2}
& =
\log^{2} \frac{ 1+s }{ p^{\ast}+s } + \mathrm{O}( n^{-1} )
\end{align}
as $T \to \infty$.
Therefore, we obtain
\begin{align}
\tilde{V}(P_{n}, W_{n})
& =
(1-p_{n}) \, (1-a_{n}) \, \bigg( \log \frac{ 1-a_{n} }{ 1-r_{n} } - I(P_{n}, W_{n}) \bigg)^{2} + (1-p_{n}) \, a_{n} \, \bigg( \log \frac{ a_{n} }{ r_{n} } - I(P_{n}, W_{n}) \bigg)^{2}
\notag \\
& \qquad \qquad {}
+ p_{n} \, (1-b_{n}) \, \bigg( \log \frac{ 1-b_{n} }{ 1-r_{n} } - I(P_{n}, W_{n}) \bigg)^{2} + (1-p_{n}) \, b_{n} \, \bigg( \log \frac{ b_{n} }{ r_{n} } - I(P_{n}, W_{n}) \bigg)^{2}
\notag \\
& =
(1-p_{n}) \, a_{n} \log^{2} \frac{ s }{ p^{\ast}+s } + p_{n} \, b_{n} \log^{2} \frac{ 1+s }{ p^{\ast}+s } + \mathrm{o}( n^{-1} )
\notag \\
& =
\frac{ A \, T }{ n } \, \bigg( (1-p^{\ast}) \, s \log^{2} \frac{ s }{ p^{\ast}+s } + p^{\ast} \, (1+s) \log^{2} \frac{ 1+s }{ p^{\ast}+s } \bigg) + \mathrm{o}( n^{-1} )
\notag \\
& =
\frac{ T \, V^{\ast} }{ n } + \mathrm{o}( n^{-1} )
\label{eq:asympt_tildeV}
\end{align}
as $T \to \infty$, implying \eqref{eq:asympt_V_direct} of \lemref{lem:unconditional_divergences}.
Finally, Equation~\eqref{eq:asympt_Xi_direct} of \lemref{lem:unconditional_divergences} can be verified in the same way as \eqref{eq:asympt_tildeV}.
This completes the proof of \lemref{lem:unconditional_divergences}.
\hfill\IEEEQEDhere

\section{Proof of \lemref{lem:delta-convention}}
\label{app:delta-convention}

We observe that
\begin{align}
\mathbb{P}\{ N(1 \mid X^{n}) > n \, \sigma \}
& \overset{\mathclap{\text{(a)}}}{=}
\mathbb{P}\{ N(1 \mid X^{n}) - \mathbb{E}[N(1 \mid X^{n})] > n \, (\sigma - p^{\ast} (1 - n^{-1/4})) \}
\notag \\
& \overset{\mathclap{\text{(b)}}}{\le}
\mathbb{P}\{ N(1 \mid X^{n}) - \mathbb{E}[N(1 \mid X^{n})] > p^{\ast} \, n^{3/4} \}
\notag \\
& \overset{\mathclap{\text{(c)}}}{\le}
\exp( - 2 \, (p^{\ast})^{2} \, \sqrt{ n } )
\label{eq:cantelli}
\end{align}
where
\begin{itemize}
\item
(a) follows from the fact that $\mathbb{E}[N(1 \mid X^{n})] = n \, p^{\ast} \,  (1-n^{-1/4})$,
\item
(b) follows by the definition of $p^{\ast}$ in \eqref{def:p_ast}, and
\item
(c) follows by Hoeffding's inequality (cf.\ \cite[Theorem~2.8]{boucheron_lugosi_massart_2013}).
\end{itemize}
This completes the proof of \lemref{lem:delta-convention} together with the hypothesis in \eqref{def:n_omegaT} that $n = \lceil T^{2} \rceil$.
\hfill\IEEEQEDhere

\section{Proof of \lemref{lem:eventEn_complement}}
\label{app:eventEn_complement}

In this proof, we employ the following lemma, which is an application of the modified logarithmic Sobolev inequality (cf.\ \cite[Chapter~6]{boucheron_lugosi_massart_2013}).

\begin{lemma}[{\cite[Theorem~6.12]{boucheron_lugosi_massart_2013}}]
\label{lem:sobolev}
Let $Z_{1}, \dots, Z_{n}$ be independent $\mathcal{A}$-valued r.v.'s.
Consider a Borel-measurable mapping $f : \mathcal{A}^{n} \to \mathbb{R}$ satisfying the self-bounding property: for each $i = 1, \dots, n$, there exists a Borel-measurable mapping $f_{i} : \mathcal{A}^{n-1} \to \mathbb{R}$ such that
\begin{align}
0
\le
f(z_{1}, \dots, z_{n}) - f_{i}(z_{1}, \dots, z_{i-1}, z_{i+1}, \dots, z_{n})
\le
1
\end{align}
and
\begin{align}
\sum_{i = 1}^{n} \Big( f(z_{1}, \dots, z_{n}) - f_{i}(z_{1}, \dots, z_{i-1}, z_{i+1}, \dots, z_{n}) \Big)
\le
f(z_{1}, \dots, z_{n})
\end{align}
for every $(z_{1}, \dots, z_{n}) \in \mathcal{A}^{n}$.
Then, it holds that
\begin{align}
\mathbb{P}\Big\{ f(Z_{1}, \dots, Z_{n}) \le \mathbb{E}[ f(Z_{1}, \dots, Z_{n}) ] - t \Big\}
\le
\exp\bigg( - \frac{ t^{2} }{ 2 \, \mathbb{E}[ f(Z_{1}, \dots, Z_{n}) ] } \bigg)
\end{align}
for every $0 < t \le \mathbb{E}[ f(Z_{1}, \dots, Z_{n}) ]$.
\end{lemma}

The following example shows a special case of \lemref{lem:sobolev}.

\begin{example}
If $Z_{1}, \dots, Z_{n}$ are independent Bernoulli r.v.'s, then the mapping
\begin{align}
f(Z_{1}, \dots, Z_{n})
=
\sum_{i = 1}^{n} Z_{i} ,
\end{align}
which follows a binomial distribution, satisfies the self-bounding property.
\end{example}

Recall that $n = \lceil T^{2} \rceil$ (see \eqref{def:n_omegaT}).
By the asymptotic equivalence
\begin{align}
n \, r_{n}
& \sim
(p^{\ast}+s) \, A \, T
\qquad (\mathrm{as} \ T \to \infty) ,
\label{eq:asympt_rn}
\end{align}
it follows from the choice of constant $0 < \kappa < 1$ that there exists a $T_{0} = T_{0}(\kappa, \lambda_{0}, A, \sigma) > 0$ satisfying
\begin{align}
\kappa \, (p^{\ast}+s) \, A \, T
\le
n \, r_{n}
\le
(1+\kappa) \, (p^{\ast}+s) \, A \, T
\label{eq:T0}
\end{align}
for every $T \ge T_{0}$.
Now, we observe that
\begin{align}
\mathbb{P}( \mathcal{E}_{T}^{\complement} )
& =
\mathbb{P}\bigg\{ \sum_{i=1}^{n} Y_{n, i} \le \mathbb{E}\bigg[ \sum_{i=1}^{n} Y_{n, i} \bigg] - \kappa \, (p^{\ast}+s) \, A \, T \bigg\}
\notag \\
& \overset{\mathclap{\text{(a)}}}{\le}
\exp \bigg( - \frac{ \kappa^{2} \, (p^{\ast}+s)^{2} \, A^{2} \, T^{2} }{ 2 \, n \, r_{n} } \bigg)
\notag \\
& \overset{\mathclap{\text{(b)}}}{\le}
\exp \bigg( - \frac{ \kappa^{2} \, (p^{\ast}+s) \, A \, T }{ 2 \, (1+\kappa) } \bigg)
\notag \\
& \overset{\mathclap{\text{(c)}}}{\le}
\exp( - K_{0} \, T ) ,
\end{align}
where
\begin{itemize}
\item
(a) holds for every $T \ge T_{0}$, because it follows from \eqref{eq:expectation_1Y}, the left-hand inequality of \eqref{eq:T0}, and \lemref{lem:sobolev},
\item
(b) holds for every $T \ge T_{0}$, because of the right-hand inequality of \eqref{eq:T0}, and
\item
(c) follows by choosing the constant
\begin{align}
K_{0}
=
K_{0}(\kappa, \lambda_{0}, A, \sigma)
=
\frac{ \kappa^{2} \, (p^{\ast}+s) \, A }{ 2 \, (1+\kappa) } .
\end{align}
\end{itemize}
This completes the proof of \lemref{lem:eventEn_complement}.
\hfill\IEEEQEDhere

\begin{remark}
\label{rem:sobolev}
While \lemref{lem:delta-convention} is proved using Hoeffding's inequality, the same inequality cannot be used to show \lemref{lem:eventEn_complement}.
This is because the Bernoulli parameter $r_{n}$ given in \eqref{def:output_rn_direct} approaches to zero as $T$ goes to infinity, but Hoeffding's inequality is independent of (the rate of decay of) $r_{n}$.
We have avoided this issue via the modified logarithmic Sobolev inequality as stated in \lemref{lem:sobolev}.
\end{remark}

\section{Proof of \lemref{lem:tight_3rd-order_term}}
\label{app:tight_3rd-order_term}

For each $n \ge 1$, let $B_{n, 1}, B_{n, 2}, \dots, B_{n, n}$ be i.i.d.\ Bernoulli r.v.'s with parameter
\begin{align}
\mathbb{P}\{ B_{n, i} = 1 \}
=
\frac{ p_{n} \, b_{n} }{ r_{n} }
\qquad \mathrm{for} \ i = 1, \dots, n .
\end{align}
For short, we write $B_{n}^{m} = (B_{n, 1}, \dots, B_{n, m})$.
\lemref{lem:eventEn_complement} tells us that there exists a $T_{0} > 0$ such that $\mathbb{E}[ N(1 \mid Y^{n}) ] > \kappa \, (p^{\ast}+s) \, A \, T$ for every $T \ge T_{0}$.
With probability one, for any reals $\alpha$ and $\beta$ and for all $T \ge T_{0}$, we have
\begin{align}
&
\bvec{1}_{\mathcal{E}_{T}} \, \mathbb{P}\bigg\{ \alpha \le \log \frac{ W_{n}^{n}(Y^{n} \mid X^{n}) }{ (P_{n}W_{n})^{n}( Y^{n} ) } \le \alpha + \beta \ \bigg| \ Y^{n} \bigg\}
\notag \\
& =
\bvec{1}_{\mathcal{E}_{T}} \sum_{\bvec{x} \in \{ 0, 1 \}^{n}} \frac{ P_{n}^{n}( \bvec{x} ) \, W_{n}^{n}(Y^{n} \mid \bvec{x}) }{ P_{n}W_{n}( Y^{n} )  } \, \bvec{1}\bigg\{ \alpha \le \log \frac{ W_{n}^{n}(Y^{n} \mid \bvec{x}) }{ (P_{n}W_{n})^{n}( Y^{n} ) } \le \alpha + \beta \bigg\}
\notag \\
& =
\bvec{1}_{\mathcal{E}_{T}} \sum_{\bvec{x} \in \{ 0, 1 \}^{n}} \bigg( \frac{ (1-p_{n}) \, (1-a_{n}) }{ 1-r_{n} } \bigg)^{N(0, 0 \mid \bvec{x}, Y^{n})} \, \bigg( \frac{ (1-p_{n}) \, a_{n} }{ r_{n} } \bigg)^{N(0, 1 \mid \bvec{x}, Y^{n})} \, \bigg( \frac{ p_{n} \, (1-b_{n}) }{ 1-r_{n} } \bigg)^{N(1, 0 \mid \bvec{x}, Y^{n})} \, \bigg( \frac{ p_{n} \, b_{n} }{ r_{n} } \bigg)^{N(1, 1 \mid \bvec{x}, Y^{n})}
\notag \\
& \qquad \qquad \quad
{} \times \bvec{1}\Bigg\{ \alpha \le \log \Bigg( \bigg( \frac{ 1-a_{n} }{ 1-r_{n} } \bigg)^{N(0, 0 \mid \bvec{x}, Y^{n})} \, \bigg( \frac{ a_{n} }{ r_{n} } \bigg)^{N(0, 1 \mid \bvec{x}, Y^{n})} \, \bigg( \frac{ 1-b_{n} }{ 1-r_{n} } \bigg)^{N(1, 0 \mid \bvec{x}, Y^{n})} \, \bigg( \frac{ b_{n} }{ r_{n} } \bigg)^{N(1, 1 \mid \bvec{x}, Y^{n})} \Bigg) \le \alpha + \beta \Bigg\}
\notag \\
& =
\bvec{1}_{\mathcal{E}_{T}} \sum_{k_{0} = 0}^{N(0 \mid Y^{n})} \sum_{k_{1} = 0}^{N(1 \mid Y^{n})} \binom{N(0 \mid Y^{n})}{ k_{0} } \, \binom{N(1 \mid Y^{n})}{ k_{1} } \, \bigg( \frac{ (1-p_{n}) \, (1-a_{n}) }{ 1-r_{n} } \bigg)^{N(0 \mid Y^{n}) - k_{0}} \, \bigg( \frac{ (1-p_{n}) \, a_{n} }{ r_{n} } \bigg)^{N(1 \mid Y^{n}) - k_{1}} \, \bigg( \frac{ p_{n} \, (1-b_{n}) }{ 1-r_{n} } \bigg)^{k_{0}} \, \bigg( \frac{ p_{n} \, b_{n} }{ r_{n} } \bigg)^{k_{1}}
\notag \\
& \qquad \qquad \qquad \qquad
{} \times \bvec{1}\Bigg\{ \alpha \le \log \Bigg( \bigg( \frac{ 1-a_{n} }{ 1-r_{n} } \bigg)^{N(0 \mid Y^{n}) - k_{0}} \, \bigg( \frac{ a_{n} }{ r_{n} } \bigg)^{N(1 \mid Y^{n}) - k_{1}} \, \bigg( \frac{ 1-b_{n} }{ 1-r_{n} } \bigg)^{k_{0}} \, \bigg( \frac{ b_{n} }{ r_{n} } \bigg)^{k_{1}} \Bigg) \le \alpha + \beta \Bigg\}
\notag \\
& =
\bvec{1}_{\mathcal{E}_{T}} \sum_{k_{0} = 0}^{N(0 \mid Y^{n})} \sum_{k_{1} = 0}^{N(1 \mid Y^{n})} \binom{N(0 \mid Y^{n})}{ k_{0} } \, \binom{N(1 \mid Y^{n})}{ k_{1} } \, \bigg( \frac{ (1-p_{n}) \, (1-a_{n}) }{ 1-r_{n} } \bigg)^{N(0 \mid Y^{n}) - k_{0}} \, \bigg( \frac{ (1-p_{n}) \, a_{n} }{ r_{n} } \bigg)^{N(1 \mid Y^{n}) - k_{1}} \, \bigg( \frac{ p_{n} \, (1-b_{n}) }{ 1-r_{n} } \bigg)^{k_{0}} \, \bigg( \frac{ p_{n} \, b_{n} }{ r_{n} } \bigg)^{k_{1}}
\notag \\
& \qquad \qquad \qquad \qquad
{} \times \bvec{1}\Bigg\{ \alpha \le N(0 \mid Y^{n}) \log \frac{ 1-a_{n} }{ 1-r_{n} } + N(1 \mid Y^{n}) \log \frac{ a_{n} }{ r_{n} } + k_{0} \log \frac{ 1-b_{n} }{ 1-a_{n} } + k_{1} \log \frac{ b_{n} }{ a_{n} } \le \alpha + \beta \Bigg\}
\notag \\
& \overset{\mathclap{\text{(a)}}}{=}
\bvec{1}_{\mathcal{E}_{T}} \sum_{k_{0} = 0}^{N(0 \mid Y^{n})} \binom{N(0 \mid Y^{n})}{ k_{0} } \, \bigg( \frac{ p_{n} \, (1-b_{n}) }{ 1-r_{n} } \bigg)^{k_{0}} \, \bigg( \frac{ (1-p_{n}) \, (1-a_{n}) }{ 1-r_{n} } \bigg)^{N(0 \mid Y^{n}) - k_{0}}
\notag \\
& \qquad \qquad
{} \times \sum_{k_{1} = 0}^{N(1 \mid Y^{n})} \binom{N(1 \mid Y^{n})}{ k_{1} } \, \bigg( \frac{ p_{n} \, b_{n} }{ r_{n} } \bigg)^{k_{1}} \, \bigg( \frac{ (1-p_{n}) \, a_{n} }{ r_{n} } \bigg)^{N(1 \mid Y^{n}) - k_{1}} \, \bvec{1}\Bigg\{ \alpha(n, k_{0}, Y^{n}) \le k_{1} \log \frac{ b_{n} }{ a_{n} } \le \alpha(n, k_{0}, Y^{n}) + \beta \Bigg\}
\notag \\
& \overset{\mathclap{\text{(b)}}}{=}
\bvec{1}_{\mathcal{E}_{T}} \sum_{k_{0} = 0}^{N(0 \mid Y^{n})} \binom{N(0 \mid Y^{n})}{ k_{0} } \, \bigg( \frac{ p_{n} \, (1-b_{n}) }{ 1-r_{n} } \bigg)^{k_{0}} \, \bigg( \frac{ (1-p_{n}) \, (1-a_{n}) }{ 1-r_{n} } \bigg)^{N(0 \mid Y^{n}) - k_{0}} \, \mathbb{P}\bigg\{ \alpha(n, k_{0}, Y^{n}) \le N(1 \mid B_{n}^{N(1 \mid Y^{n})}) \log \frac{ b_{n} }{ a_{n} } \le \alpha(n, k_{0}, Y^{n}) + \beta \bigg\}
\notag \\
& \overset{\mathclap{\text{(c)}}}{\le}
\bvec{1}_{\mathcal{E}_{T}} \sum_{k_{0} = 0}^{N(0 \mid Y^{n})} \binom{N(0 \mid Y^{n})}{ k_{0} } \, \bigg( \frac{ p_{n} \, (1-b_{n}) }{ 1-r_{n} } \bigg)^{k_{0}} \, \bigg( \frac{ (1-p_{n}) \, (1-a_{n}) }{ 1-r_{n} } \bigg)^{N(0 \mid Y^{n}) - k_{0}} \, \Bigg( \Phi\bigg( \frac{ \alpha(n, k_{0}, Y^{n}) + \beta }{ \log(b_{n}/a_{n}) } \bigg) - \Phi\bigg( \frac{ \alpha(n, k_{0}, Y^{n}) }{ \log (b_{n}/a_{n}) } \bigg) +  \frac{ 12 \, \xi_{n} }{ \sigma_{n}^{2} } \Bigg) \nonumber\\*
&\qquad\qquad\times\frac{ 1 }{ \sqrt{ N(1 \mid Y^{n}) \, \sigma_{n}^{2} } }
\notag \\
& \le
\bvec{1}_{\mathcal{E}_{T}} \underbrace{ \sum_{k_{0} = 0}^{N(0 \mid Y^{n})} \binom{N(0 \mid Y^{n})}{ k_{0} } \, \bigg( \frac{ p_{n} \, (1-b_{n}) }{ 1-r_{n} } \bigg)^{k_{0}} \, \bigg( \frac{ (1-p_{n}) \, (1-a_{n}) }{ 1-r_{n} } \bigg)^{N(0 \mid Y^{n}) - k_{0}} }_{= 1} \, \Bigg( \frac{ \beta }{ (\log(b_{n}/a_{n})) \sqrt{ 2 \, \pi } } +  \frac{ 12 \, \xi_{n} }{ \sigma_{n}^{2} } \Bigg) \frac{ 1 }{ \sigma_{n} \, \sqrt{ N(1 \mid Y^{n}) } }
\notag \\
& =
\bvec{1}_{\mathcal{E}_{T}} \, \Bigg( \frac{ \beta }{ (\log(b_{n}/a_{n})) \sqrt{ 2 \, \pi } } +  \frac{ 12 \, \xi_{n} }{ \sigma_{n}^{2} } \Bigg) \frac{ 1 }{ \sigma_{n} \, \sqrt{ N(1 \mid Y^{n}) } }
\notag \\
& \overset{\mathclap{\text{(d)}}}{\le}
\bvec{1}_{\mathcal{E}_{T}} \, \Bigg( \frac{ \beta }{ (\log(b_{n}/a_{n})) \sqrt{ 2 \, \pi } } +  \frac{ 12 \, \xi_{n} }{ \sigma_{n}^{2} } \Bigg) \frac{ 1 }{ \sigma_{n} \, \sqrt{ \mathbb{E}[ N(1 \mid Y^{n}) ] - \kappa \, (p^{\ast}+s) \, A \, T } }
\notag \\
& \le
\Bigg( \frac{ \beta }{ (\log(b_{n}/a_{n})) \sqrt{ 2 \, \pi } } +  \frac{ 12 \, \xi_{n} }{ \sigma_{n}^{2} } \Bigg) \frac{ 1 }{ \sigma_{n} \, \sqrt{ \mathbb{E}[ N(1 \mid Y^{n}) ] - \kappa \, (p^{\ast}+s) \, A \, T } }
\notag \\
& \overset{\mathclap{\text{(e)}}}{=}
\Bigg( \frac{ \beta }{ (\log(b_{n}/a_{n})) \sqrt{ 2 \, \pi } } +  \frac{ 12 \, \xi_{n} }{ \sigma_{n}^{2} } \Bigg) \frac{ 1 }{ \sigma_{n} \, \sqrt{ n \, r_{n} - \kappa \, (p^{\ast}+s) \, A \, T } }
\notag \\
& =
\Bigg( \frac{ \beta }{ (\log(b_{n}/a_{n})) \sqrt{ 2 \, \pi } } +  \frac{ 12 \, \xi_{n} }{ \sigma_{n}^{2} } \Bigg) \, \frac{ 1 }{ \sigma_{n} \, \sqrt{ T } } \, \sqrt{ \frac{ T }{ n \, r_{n} - \kappa \, (p^{\ast}+s) \, A \, T } } ,
\label{eq:interval_Berry-Esseen}
\end{align}
where
\begin{itemize}
\item
(a) follows by defining the r.v.\ $\alpha(n, k_{0}, Y^{n})$ so that
\begin{align}
\alpha(n, k_{0}, Y^{n})
& \coloneqq
\alpha - N(0 \mid Y^{n}) \log \frac{ 1-a_{n} }{ 1-r_{n} } - N(1 \mid Y^{n}) \log \frac{ a_{n} }{ r_{n} } - k_{0} \log \frac{ 1-b_{n} }{ 1-a_{n} } ,
\end{align}
\item
(b) follows from the fact that the r.v.\
\begin{align}
N(1 \mid B_{n}^{N(1 \mid Y^{n})})
=
\sum_{i = 1}^{N(1 \mid Y^{n})} B_{n, i}
\end{align}
follows the binomial distribution with parameters $N(1 \mid Y^{n})$ and $p_{n} \, b_{n} / r_{n}$,
\item
(c) follows from the Berry--Esseen theorem with the identities:
\begin{align}
\sigma_{n}^{2}
& \coloneqq
\mathbb{E}[ (B_{n, 1} - \mathbb{E}[ B_{n, 1} ])^{2} ]
\notag \\
& \: =
\frac{ p_{n} \, (1-p_{n}) \, a_{n} \, b_{n} }{ r_{n}^{2} } ,
\\
\xi_{n}
& \coloneqq
\mathbb{E}[ |B_{n, 1} - \mathbb{E}[ B_{n, 1} ]|^{3} ]
\notag \\
& \: =
\frac{ (1-p_{n}) \, a_{n} }{ r_{n} } \, \bigg( \frac{ p_{n} \, b_{n} }{ r_{n} } \bigg)^{3} + \frac{ p_{n} \, b_{n} }{ r_{n} } \, \bigg( 1 - \frac{ p_{n} \, b_{n} }{ r_{n} } \bigg)^{3} ,
\end{align}
\item
(d) follows by the definition of $\mathcal{E}_{T}$ in \eqref{def:event_n}, and
\item
(e) follows from \eqref{eq:expectation_1Y}.
\end{itemize}
Recall that $n = \omega( T )$ as $T \to \infty$.
By the asymptotic equivalences
\begin{align}
n \, a_{n}
& \sim
s \, A \, T ,
\\
n \, b_{n}
& \sim
(1+s) \, A \, T ,
\\
n \, r_{n}
& \sim
(p^{\ast}+s) \, A \, T
\end{align}
as $T \to \infty$, we see that
\begin{align}
\lim_{T \to \infty} \log \frac{ b_{n} }{ a_{n} }
& =
\log \frac{ 1+s }{ s } ,
\\
\lim_{T \to \infty} \sigma_{n}^{2}
& =
\frac{ p^{\ast} \, (1-p^{\ast}) \, s \, (1+s) }{ (p^{\ast}+s)^{2} } ,
\\
\lim_{T \to \infty} \xi_{n}
& =
\frac{ (1-p^{\ast}) \, s }{ p^{\ast}+s } \, \bigg( \frac{ p^{\ast} \, (1+s) }{ p^{\ast}+s } \bigg)^{3} + \frac{ p^{\ast} \, (1+s) }{ p^{\ast}+s } \, \bigg( 1 - \frac{ p^{\ast} \, (1+s) }{ p^{\ast}+s } \bigg)^{3} ,
\end{align}
and
\begin{align}
\lim_{T \to \infty} \sqrt{ \frac{ T }{ n \, r_{n} - \kappa \, (p^{\ast}+s) \, A \, T } }
& =
\frac{ 1 }{ \sqrt{ (1 - \kappa) \, (p^{\ast}+s) \, A } } .
\end{align}
Therefore, it follows from \eqref{eq:interval_Berry-Esseen} that there exist $K_{1} = K_{1}(\kappa, \lambda_{0}, A, \sigma) > 0$ and $T_{1} = T_{1}(\kappa, \lambda_{0}, A, \sigma) > 0$ satisfying
\begin{align}
\bvec{1}_{\mathcal{E}_{T}} \, \mathbb{P}\bigg\{ \alpha \le \log \frac{ W_{n}^{n}(Y^{n} \mid X^{n}) }{ (P_{n}W_{n})^{n}( Y^{n} ) } \le \alpha + \beta \ \bigg| \ Y^{n} \bigg\}
\le
\frac{ K_{1} }{ 2 \, \sqrt{ T } }
\label{eq:one_over_sqrtT}
\end{align}
almost surely for every reals $\alpha$ and $\beta$ and every $T \ge T_{1}$.
Now, as in \cite[Equation~(474)]{polyanskiy_poor_verdu_2010}, it follows from \eqref{eq:one_over_sqrtT} that
\begin{align}
&
\bvec{1}_{\mathcal{E}_{T}} \, \mathbb{E}\Big[ \mathrm{e}^{- \iota_{n}(X^{n} \wedge Y^{n})} \, \bvec{1}\{ \iota_{n}(X^{n} \wedge Y^{n}) \ge \gamma \} \ \Big| \ Y^{n} \Big]\nonumber\\*
& \quad \le
\sum_{l = 0}^{\infty} \exp(- \gamma - l \log 2) \, \bvec{1}_{\mathcal{E}_{T}} \, \mathbb{P}\bigg\{ \gamma + l \log 2 \le \log \frac{ W_{n}^{n}(Y^{n} \mid X^{n}) }{ (P_{n}W_{n})^{n}( Y^{n} ) } < \gamma + (1+l) \log 2 \ \bigg| \ Y^{n} \bigg\}
\notag \\
& \quad \le
\frac{ K_{1} \, \exp( - \gamma ) }{ 2 \, \sqrt{ T } } \sum_{l = 0}^{\infty} 2^{-l}
\notag \\
& \quad =
\frac{ K_{1} \, \exp( - \gamma ) }{ \sqrt{ T } }
\end{align}
almost surely for every $T \ge T_{1}$.
This completes the proof of \lemref{lem:tight_3rd-order_term}.
\hfill\IEEEQEDhere

\section{Proof of \lemref{lem:final_sqrt}}
\label{app:final_sqrt}

It follows from \cite[Lemma~47]{polyanskiy_poor_verdu_2010} (see also \cite[Theorem~1.7]{tan_2014}) that
\begin{align}
\mathbb{E} \Big[ \bvec{1}\{ \iota_{n}(X^{n} \wedge Y^{n}) > \gamma \} \, \exp( - \iota_{n}(X^{n} \wedge Y^{n}) ) \Big]
\le
2 \bigg( \frac{ \log 2 }{ \sqrt{ 2 \, \pi } } + \frac{ \tilde{\Xi}(P_{n}, W_{n}) }{ \tilde{V}(P_{n}, W_{n}) } \bigg) \frac{ \exp( - \gamma ) }{ \sqrt{ n \, \tilde{V}(P_{n}, W_{n}) } } ,
\label{eq:indicator_exp_bound_usage}
\end{align}
where note that
\begin{align}
\iota_{n}(X^{n} \wedge Y^{n})
& =
\sum_{i = 1}^{n} \log \frac{ W_{n}(Y_{n, i} \mid X_{n, i}) }{ P_{n}W_{n}( Y_{n, i} ) }
\end{align}
and
\begin{align}
\mathbb{E}\bigg[ \log \frac{ W_{n}(Y_{n, 1} \mid X_{n, 1}) }{ P_{n}W_{n}( Y_{n, 1} ) } \bigg]
& =
I(P_{n}, W_{n}) ,
\\
\mathbb{E}\bigg[ \bigg( \log \frac{ W_{n}(Y_{n, 1} \mid X_{n, 1}) }{ P_{n}W_{n}( Y_{n, 1} ) } - I(P_{n}, W_{n}) \bigg)^{2} \bigg]
& =
\tilde{V}(P_{n}, W_{n}) ,
\\
\mathbb{E}\bigg[ \bigg| \log \frac{ W_{n}(Y_{n, 1} \mid X_{n, 1}) }{ P_{n}W_{n}( Y_{n, 1} ) } - I(P_{n}, W_{n}) \bigg|^{3} \bigg]
& =
\tilde{\Xi}(P_{n}, W_{n}) .
\end{align}
Therefore, \lemref{lem:unconditional_divergences} yields \lemref{lem:final_sqrt}.
\hfill\IEEEQEDhere

\section{Proof of \lemref{lem:Berry-Esseen_unconditional}}
\label{app:Berry-Esseen_unconditional}

It follows from the Berry--Esseen theorem (cf.\ \cite[Theorem~1.6]{tan_2014}) that
\begin{align}
\mathbb{P}\bigg\{ \iota_{n}(X^{n} \wedge Y^{n}) \le \gamma \bigg\}
\le
\Phi\left( \frac{ \gamma - n \, I(P_{n}, W_{n}) }{ \sqrt{ n \, \tilde{V}(P_{n}, W_{n}) } } \right) + \frac{ 6 \, \tilde{\Xi}(P_{n}, W_{n}) }{ \sqrt{ n \, \tilde{V}(P_{n}, W_{n})^{3} } } .
\end{align}
The existence of the constant $K_{3} = K_{3}(\lambda_{0}, A, \sigma) > 0$ can be verified by \lemref{lem:unconditional_divergences}.
This completes the proof of \lemref{lem:Berry-Esseen_unconditional}.
\hfill\IEEEQEDhere

\bibliographystyle{IEEEtran}
\bibliography{IEEEabrv,mybib}

\begin{IEEEbiographynophoto}{Yuta Sakai}
(S'16-M'19) was born in Japan in 1992.
He is currently a Research Fellow in the Department of Electrical and Computer Engineering at the National University of Singapore (NUS).
He received the B.E.\ and M.E.\ degrees in the Department of Information Science from the University of Fukui in 2014 and 2016, respectively, and the Ph.D.\ degree in the Advanced Interdisciplinary Science and Technology from the University of Fukui in 2018.
His research interests include information theory and coding theory, 
\end{IEEEbiographynophoto}

\begin{IEEEbiographynophoto}{Vincent Y.\ F.\ Tan}
(S'07-M'11-SM'15) was born in Singapore in 1981. He is currently a Dean's Chair Associate Professor in the Department of Electrical and Computer Engineering and the Department of Mathematics at the National University of Singapore (NUS).
He received the B.A.\ and M.Eng.\ degrees in Electrical and Information Sciences from Cambridge University in 2005 and the Ph.D.\ degree in Electrical Engineering and Computer Science (EECS) from the Massachusetts Institute of Technology (MIT)  in 2011.  His research interests include information theory, machine learning, and statistical signal processing.

Dr.\ Tan received the MIT EECS Jin-Au Kong outstanding doctoral thesis prize in 2011, the NUS Young Investigator Award in 2014,  the Singapore National Research Foundation (NRF) Fellowship (Class of 2018) and the NUS Young Researcher Award in 2019. He was also an IEEE Information Theory Society Distinguished Lecturer for 2018/9. He is currently serving as an Associate Editor of the IEEE Transactions on Signal Processing and an Associate Editor of Machine Learning for the IEEE Transactions on Information Theory.
\end{IEEEbiographynophoto}

\begin{IEEEbiographynophoto}{Mladen Kova\v{c}evi\'{c}}
was born in Sarajevo, Bosnia and Herzegovina, in 1984.
He received the Dipl.Ing.\ (2007) and Ph.D.\ (2014) degrees in Electrical and Computer Engineering from the University of Novi Sad, Serbia, where he also currently works as an Assistant Professor.
In the period from Oct.\ 2015 to May 2018 he worked at the National University of Singapore, as a Research Fellow in the group of Vincent Y.~F.~Tan. His research interests include information theory, coding theory, and discrete mathematics.
\end{IEEEbiographynophoto}

\end{document}